\documentclass[applsci,article,accept,moreauthors,pdftex]{Definitions/mdpi} 

\firstpage{1} 
\pubvolume{12}
\issuenum{2}
\articlenumber{0705}
\pubyear{2022}
\copyrightyear{2022}
\externaleditor{Academic Editors: Roberta Sparvoli and Matteo Martucci}
\datereceived{8 November 2021} 
\dateaccepted{23 December 2021} 
\datepublished{11 January 2022} 
\hreflink{https://doi.org/10.3390/\linebreak app12020705} % If needed use \linebreak
\Title{Measurement of Energy Spectrum and Elemental Composition of PeV Cosmic Rays: Open Problems and Prospects}

\TitleCitation{Measurement of Energy Spectrum and Elemental Composition of PeV Cosmic Rays: Open Problems and Prospects}

\Author{{Giuseppe Di Sciascio} %Please carefully check the accuracy of names and affiliations.
\orcidA{}}

\AuthorNames{Giuseppe Di Sciascio}

\AuthorCitation{Di Sciascio, G.}
\address[1]{%
{INFN---Roma Tor Vergata, Department of Physics, University of Roma Tor Vergata, Viale della Ricerca Scientifica 1, I-00133 Roma, Italy}; disciascio@roma2.infn.it}
\abstract{Cosmic rays represent one of the most important energy transformation processes of the universe.
They bring information about the surrounding universe, our galaxy, and very probably also the extragalactic space, at least at the highest observed energies. 
More than one century after their discovery, we have no definitive models yet about the origin, acceleration and propagation processes of the radiation.
The main reason is that there are still significant discrepancies among the results obtained by different experiments located at ground level, probably due to unknown systematic uncertainties affecting the measurements.
In this document, we will focus on the detection of galactic cosmic rays from ground with air shower arrays up to 10$^{18}$ eV. 
The aim of this paper is to discuss the conflicting results in the 10$^{15}$ eV energy range and the perspectives to clarify the origin of the so-called \emph{`knee'} in the all-particle energy spectrum, crucial to give a solid basis for models up to the end of the cosmic ray spectrum.
We will provide elements useful to understand the basic techniques used in reconstructing primary particle characteristics (energy, mass, and arrival direction) from the ground, and to show why indirect measurements are difficult and results are still conflicting. 
}


\begin{document}
%%%%%%%%%%%%%%%%%%%%%%%%%%%%%%%%%%%%%%%%%%
%\setcounter{section}{-1} %% Remove this when starting to work on the template.
\section{Introduction}

Cosmic rays (CRs) are the most outstanding example of accelerated particles and represent about 1\% of the total mass of the universe \cite{battistonigrillo1996}. The riddle of the origin of this radiation has been unsolved for more than a century.
The study of CRs is based on two complementary approaches \cite{disciascio2019}:
\begin{enumerate}
\item[(1)] Measurement of energy spectrum, elemental composition and anisotropy in the CR arrival direction distribution, the three basic parameters crucial for understanding the origin, acceleration, and propagation of  radiation.
\item[(2)] Search of their sources through the observation of neutral radiation (photons and neutrinos), which points back to the emitting sources not being affected by the magnetic fields, in a multi-messenger approach. We note that, however, photons and neutrinos do not necessarily point back to their sources (see, for example, the the Ref. \cite{stanev2014}).
\end{enumerate}%Is the italics necessary?
%We note that, however, photons and neutrinos do not necessarily point back to their sources. As an example, photons and neutrinos can be produced in molecular clouds on the way in the galaxy or during intergalactic propagation for cosmogenic particles \cite{stanev2014}.

\begin{figure}[H]
\includegraphics[width=1\textwidth]{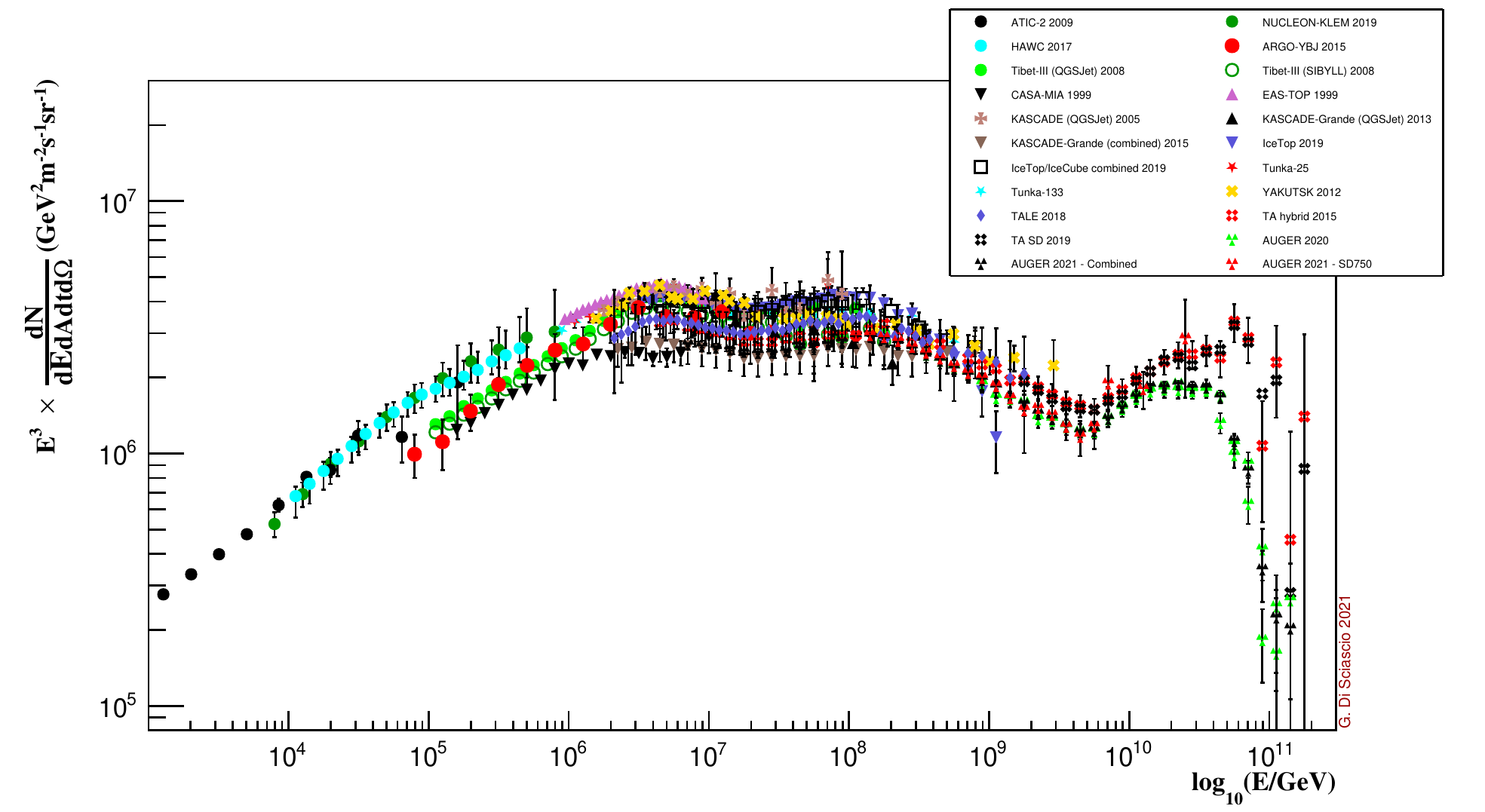}
\caption{All–particle energy spectrum of primary cosmic rays, the flux is multiplied by E$^3$. Results from direct and indirect experiments updated to the year 2021 are shown.}
\label{fig:allpart_enespt}       % Give a unique label
\end{figure}

In Figure~\ref{fig:allpart_enespt}, the primary CR all-particle energy spectrum (namely, the number of nuclei as a function of total energy) is shown. 
The spectrum exceeds 10$^{20}$ eV, showing a few basic characteristics \cite{disciascio2019}: 
%
%%\begin{itemize}
\begin{enumerate}[label=(\roman*)]
\item[(a)] A power-law behaviour $\sim$E$^{-2.7}$ up to the so-called \emph{``knee''}, a small downwards bend around a few PeV (1 PeV = 10$^{15}$ eV); 
\item[(b)] a power-law behaviour $\sim$E$^{-3.1}$ beyond the knee, with a downwards bend near 10$^{17}$ eV, sometimes referred to as the \emph{``second knee''}; 
\item[(c)] a transition back to a power-law $\sim$E$^{-2.7}$ (the so-called \emph{``ankle''}) around $10^{18.7}$ eV; 
\item[(d)] a cutoff, probably due to extra-galactic CR interactions with the Cosmic Microwave Background (CMB), around 10$^{19.7}$ eV (the Greisen-Zatsepin-Kuzmin effect).
\end{enumerate}
%
%%%%%%%%%%%%%%%%%%%%%%%%%%%%%%%%%%%%
%MDPI: please   change the hyphen to minus sign, -1 should be  −1
%%%%%%%%%%%%%%%%%%%%%%%%%%%%%%%%%%%%

Despite the differences in flux, emphasized by multiplying the differential spectrum by E$^3$, all the measurements of the all-particle energy spectrum are in fair agreement when taking into account the statistical, systematic and energy scale uncertainties. 
Nevertheless, uncertainties affecting flux measurements could be underestimated for a number of reasons discussed in this document. In a conservative approach, the spread of different results provides a more realistic estimate of the uncertainty.

All the observed features are believed to carry fundamental information that sheds light on the key questions of the origin, acceleration and propagation of CRs. However, from the all-particle results alone, it is not possible to understand the origin of different features. All models concerning sources, acceleration and propagation of the primary flux, differ considerably for what concerns expected elemental composition as a function of the energy. A measurement of the chemical composition is therefore crucial to disentangle between different hypotheses.

The main structure is the \emph{``knee''} observed for the first time by R.W. Williams in 1948 in the experiment which first located individual shower cores from symmetry of the fired detectors \cite{williams1948,linsley1983}.
The knee as a feature connected to the end of the Galactic CR flux was first suggested in 1959 by Kulikov and Khristiansen \cite{khristiansen1959}. They speculated that particles above 10$^{16}$ eV may have a \emph{``metagalactic origin''}. Consequently, the observed spectrum is a superposition of the spectra of particles of galactic and metagalactic origin.
In 1962, Miura and Hasegawa \cite{miura1962} reported the first observation of two spectral kinks (in both N$_e$ and N$_\mu$ spectra) correlating them to a steepening of the primary energy spectrum.

All experiments observed the knee at about 4 ~$\times$~ 10$^{15}$ eV but a general consensus about the chemical component responsible for such a feature does not exist yet because experimental results are still conflicting, as will be discussed in Section~\ref{sec:composition}. 
Determining elemental composition in the knee energy region is crucial to understand where Galactic CR spectrum ends and to give a solid basis to CR models up to the highest observed energies.
The maximum energy at which the various nuclei are accelerated should be subject to a rigidity cutoff, as proposed originally by Peters \cite{peters1961}. Protons will cutoff first, followed by other nuclei according to the relation
\begin{equation}
E_\text{max}(Z) = Z\times E_\text{max}(Z=1)
\end{equation}

If the dominant primary mass of the knee is light (protons and helium), then, according to this scheme, the Galactic CR spectrum is expected to end around 10$^{17}$ eV with iron. 
The sum of the fluxes of all elements, with their individual knees at energies proportional to the nuclear charge, makes up the CR all-particle spectrum shown in Figure~\ref{fig:allpart_enespt}.
With increasing energies, not only does the spectrum become steeper due to such cutoffs, but also heavier.
In this scenario, the knee would represent the end of the spectrum of CR accelerated by SNRs in the galaxy.

Indeed, it is widely believed that the bulk of CRs up to about 10$^{17}$ eV are galactic, produced and accelerated by the shock waves of SuperNova Remnants (SNR) expanding shells \cite{drury12}, and that the transition to extra-galactic CRs occurs somewhere between 10$^{17}$ and 10$^{19}$ eV.
The experimental results, however, do not demonstrate the capability of SNRs to produce the power needed to sustain the population of galactic CRs and to accelerate particles up to the knee, and beyond. 
Indeed, to accelerate protons up to the PeV energy domain, a significant amplification of the magnetic field at the shock is required, but this process is problematic \cite{gabici16}.

Unlike neutrinos that are produced only in hadronic interactions of CRs, the question whether the observed $\gamma$-rays are produced by the decay of $\pi^0$ from CR interactions (\emph{`hadronic'} mechanism), or by a population of relativistic electrons via Inverse Compton scattering or bremsstrahlung (\emph{`leptonic'} mechanism), still needs a conclusive answer.
In a hadronic interaction, the secondary photons have, on average, an energy factor of 10, lower than the primary proton. 
Therefore, the quest for CR sources to be able to accelerate particles up to the PeV range in a multi-messenger approach requires the observation of the $\gamma$-ray sky above 100 TeV. 
However, the first results reported by the LHAASO experiment~\cite{cao2021nat,cao2021scie}, that is, the observation of a number of gamma sources emitting photons beyond 500 TeV, show that SNRs are likely not the main sources of PeV CRs in our galaxy. 
In fact, none of the 12 observed ultra-high energy gamma sources can be clearly described with hadronic mechanisms operating in SNRs. We note that the highest photon emission at 1.4~PeV comes from a system of massive stars in the Cygnus Region, the so-called \emph{`Cygnus Cocoon'}, a possible factory of fresh CRs, as suggested by other experiments \cite{bartoli2014,aharonian2019}.

In this note, we will focus on galactic CRs in the PeV energy range detected from ground with air shower arrays.
This is not a place for a complete review of CR physics and models (for which we recommend, for instance, \cite{spurio,gaisser,grieder,longair,blasi,disciascio2019} and the references therein), but only to provide elements useful to understand the main techniques used in reconstructing primary particle characteristics from the ground with particle arrays, and to show why indirect measurements are difficult and the results are still conflicting. 

In the next section, we will introduce the detection techniques. In Section \ref{sec:easmodel} we will describe the main characteristics of Extensive Air Showers to understand how different observables measured by arrays are related to the properties of the primary CRs. In  \mbox{Section \ref{enemassrec}}, we will discuss the general scheme of the air shower array analysis. In  Section~\ref{sec:composition} the experimental results in the 10$^{14}$--10$^{18}$ eV energy range are summarized. The prospects for new measurements in the knee region are introduced in  Section \ref{sec6}.

\section{Detection Techniques}
We can divide the experimental methods adopted to measure fluxes and elemental composition of CRs into two categories: \emph{`direct'} and \emph{`indirect'} measurements. 
Generally speaking, for all particle types:
\begin{itemize}
\item the higher the energy, the lower the flux;
\item the lower the flux, the larger the required detector area.
\end{itemize}

The direct measurements, in principle, detect and directly identify  the primary particles with detectors outside the atmosphere (on board of stratospheric balloons or satellites), since the atmosphere behaves as a shield (see below). 
Since the CR flux rapidly decreases with increasing energy and the size of detectors is constrained by the weight that can be carried in flight, their \emph{`aperture'} (i.e., the acceptance measured in m$^2\cdot$sr) is small and determines a maximum energy (of the order of a few hundred TeV/nucleon), at which a statistically significant detection is possible. 
In fact, the number of detected events is given by the CR flux times the detection area times the total observation time.
Therefore, the detection area limits the smallest measurable flux.
In addition, the limited volume of the detectors makes  the containment of showers induced by high-energy nuclei difficult, thus limiting the energy resolution of the instruments in direct measurements.

At higher energies, the flux is so low (about 1 particle/m$^2$/year around 10$^{15}$ eV) that the only chance is to have earth-based detectors of large area, operating for long times. In that case, the atmosphere is considered as a target, and we study the primary properties in an \emph{`indirect'} way, through the measurement of secondary particles produced in the interaction of the primary particle with the nuclei of the atmosphere, the so-called \emph{`Extensive Air Shower'} (EAS).

Approaching the hundred TeV energy region, even in space-borne experiments, the energy assignment is indirect since it is generally based on the energy deposition of particles produced in the interaction of primaries in the detector itself. The reconstruction of the total energy is then obtained by comparison with some model prediction, and therefore, at least in that region, the boundary line between `direct' and `indirect' experiments is more uncertain.
In fact, important results obtained by `direct' methods are conflicting due to some still unknown systematic uncertainties probably related to the interaction model used to assign the energy. 
A neutrino energy-dependent component must be estimated via Monte Carlo simulations, an evaluation which adds some additional model dependency for 'indirect' measurements.

At the ground, the study of CRs is based on the reconstruction and interpretation of EAS observables in the different components, electromagnetic (e.m.), muonic and hadronic, Cherenkov photons, nitrogen fluorescence, radio emission.
Therefore, different detectors must be used to detect different observables.

Two different approaches are exploited:

\begin{itemize}

\item Arrays, to sample the shower tail particles reaching the ground. In High Energy Particle language, a shower array is a \emph{``Tail Catcher Sampling Calorimeter''}. The atmosphere is the absorber and the detectors at ground are the device to measure a (poor) calorimetric signal. Arrays are wide field of view detectors able to observe most of the overhead sky with a duty cycle of $\sim$100\%. Measurements are limited by large shower-to-shower fluctuations.
\item Telescopes, to detect Cherenkov photons or nitrogen fluorescence and observe the EAS longitudinal profile. The atmosphere acts as a \emph{``Homogeneous Calorimeter''}. The duty cycle is low ($\sim$10--15\%) because telescopes can be operated only during clear moonless nights and the field of view is small (a few degrees). On the contrary, pointing capability and energy resolution are excellent.
\end{itemize}%is the bold necessary?

\begin{figure}[H]
{\includegraphics[width=0.95\textwidth,clip]{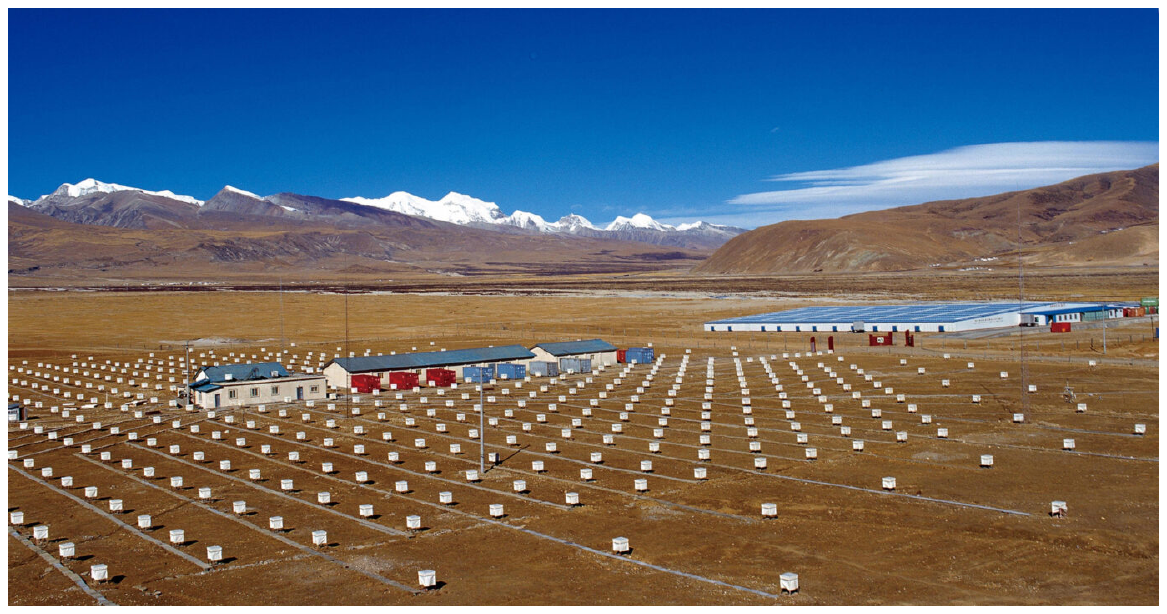} }
\caption{Example of a typical air shower array (Tibet AS$\gamma$ experiment located at the YangBaJing Cosmic Ray Observatory in Tibet (P.R. China) 4300 m asl).}
\label{fig:array}       % Give a unique label, MDPI: Please replace with a sharper image.

\end{figure}

Shower arrays are made by a large number of detectors (scintillators, Resistive Plate Chambers (RPCs) or water Cherenkov tanks, for example) distributed in a regular grid over very large areas, of an order of 10$^{4}$--10$^{5}$ m$^2$ (see Figure~\ref{fig:array}). 
The shower \emph{``size''}, the total number of charged particles, and the shower arrival direction are the two key parameters reconstructed by all arrays.
The majority of EAS arrays do not distinguish between the charged particles. From the measurement of the particle densities on the fired detectors of the array it is possible to determine the shower core position, that is, the point where the shower axis intersects the detection plane, and, via a Lateral Density Function (LDF), reconstruct the size of the shower.
The LDF is of phenomenological nature, determined via Monte Carlo simulations for the particular experimental set-up \cite{bartoli2011}.
The direction of the incoming primary particle is reconstructed with a \emph{`time of flight'} method making use of the relative times at which the individual detection units are fired by the shower front \cite{bartoli2011}.
%
%%%%%%%%%%%%%%%%%%%%%%%%%%%%%%%%%%%%%%%%%%%%%%%%%%%%%%%%%%
%%%%%%%%%%%%%%%%%%%%%%%%%%%%%%%%%%%%%%%%%%%%%%%%%%%%%%%%%%
%

On general grounds, the instrumented area A determines the rate of high energy events recorded, that is, the maximum energy via limited statistics. The grid distance $d$ determines the low energy threshold (small energy showers are lost in the gap between detectors) and the quality of the shower sampling. The particular kind of detector (scintillator, RPC, water tank) determines the detail of measurement (efficiency, resolution, energy threshold, quality) and impact on the cost per detector $C_d$. In principle, best physics requires large area $A$, small distance $d$ and high quality of the sampling. However, the cost of an array increases with $C_d\cdot A/d^2$, therefore a compromise is always needed. 
This is one of the reason why the typical total sensitive area of a classical array is less than 1\% of the total enclosed area. This results in a high degree of uncertainty in the reconstruction due to sampling fluctuations which add to the shower fluctuations.

The experiments devoted to study the PeV energy range have been operated at different altitudes, ranging from the extreme altitude (5200 m asl) of BASJE-MAS \cite{basje-mas} to the sea level of KASCADE \cite{kascade1,kascade2,kascade3}. %references 23 and 24 are not cited in the main text. Please revise.

In Tables \ref{tab:array-summary} and \ref{tab:array-summary-muon}, the characteristics of air shower arrays operated in the last two decades to study Galactic CR physics from ground are summarized. 
The atmospheric depths of the arrays, the main detectors used, the energy range investigated, the sensitive areas of e.m. and muon detectors, the instrumented areas and the coverage (i.e., the ratio between sensitive and instrumented areas) are reported.
The depth in atmosphere is crucial to fix the energy threshold, the energy resolution, the impact of shower-to-shower fluctuations, then the sensitivity to elemental composition. 
%
%%%%%%%%% Tables 1-2 %%%%%%%%%%%%%%%%%%%%%%%%%%%%%%%%%%%%%%%%%%%%
\begin{table}[H]
\setlength{\tabcolsep}{5.15mm}
\caption{\label{tab:array-summary} Characteristics of different air shower arrays.}
\footnotesize
\begin{adjustwidth}{-\extralength}{0cm}
\centering
\begin{tabular}{lllllll}
\toprule
  \multirow{2}{*}{\textbf{Experiment}} &  \multirow{2}{*}{\textbf{g/cm}\boldmath{$^2$}} & \multirow{2}{*}{\textbf{Detector}} & \boldmath{$\Delta$}\textbf{E} & \textbf{e.m.\ Sens.} & \textbf{Instr.} & \multirow{2}{*}{\textbf{Coverage}} \\
                     &                  &                &       \textbf{(eV)}       &         \textbf{Area} \textbf{(m}\boldmath{$^2$}\textbf{)}             & \textbf{Area} \textbf{(m}\boldmath{$^2$}\textbf{)} & \\
\midrule
ARGO-YBJ \cite{bartoli2011}  & 606            & RPC/hybrid  with  & $3\times 10^{11} $--$ 10^{16}$ & 6700 & 11,000 & 0.93 \\
                    &                   &   wide-FoV \v C Tel.   &                                            &          &  & (carpet)\\\midrule
BASJE-MAS \cite{basje-mas} & 550           & scint./muon & $6\times 10^{12} $--$ 3.5\times 10^{16}$ &    & $10^4$ &  \\\midrule
 TIBET AS$\gamma$ \cite{amenomori2011} & 606 &  scint./burst det.  & $5\times 10^{13} $--$ 10^{17}$ & 380 & 3.7~$\times$~10$^4$ & 10$^{-2}$ \\\midrule
 CASA-MIA \cite{casamia} & 860 & scint./muon & 10$^{14} $--$ 3.5\times 10^{16}$ & 1.6~$\times$~10$^3$ & 2.3~$\times$~10$^5$ & 7~$\times$~10$^{-3}$ \\\midrule
KASCADE \cite{kascade1} & 1020 & scint./mu/had & $2\times 10^{15} $--$ 10^{17}$ & 5~$\times$~10$^2$ & 4~$\times$~10$^4$ & 1.2~$\times$~10$^{-2}$ \\\midrule
KASCADE-& 1020 & scint./mu/had  & $10^{16} $--$ 10^{18}$ & 370 & 5~$\times$~10$^5$ & 7~$\times$~10$^{-4}$ \\
Grande \cite{kascade-grande} &  &  &  &  &  &  \\\midrule
Tunka \cite{prosin2014} & 900 & open \v C det. & 3~$\times 10^{15} $--$ 3\times 10^{18}$ & --- & 10$^6$ & --- \\\midrule
IceTop \cite{aartsen2019} & 680 & ice \v C det. & $10^{16} $--$ 10^{18}$ & 4.2~$\times$~10$^2$ & 10$^6$ & 4~$\times$~10$^{-4}$ \\\midrule
 LHAASO \cite{cao2021nat} & 600 & Water \v C& $10^{12} $--$ 10^{17}$ & 5.2~$\times$~10$^3$ & 1.3~$\times$~10$^6$ & 4~$\times$~10$^{-3}$ \\
                    &        & scint./mu/had     &                                            &          &  & \\
                    &        & wide-FoV \v C Tel.     &                                            &          &  & \\
\bottomrule
\end{tabular}
\end{adjustwidth}
\end{table}
\vspace{-12pt}
\begin{table}[H]\setlength{\tabcolsep}{3.5mm}
\caption{\label{tab:array-summary-muon} Characteristics of different muon detectors operated in some shower arrays.}
\begin{tabular}{ccccc}
\toprule
 \multirow{2}{*}{\textbf{Experiment}} &\textbf{Altitude} & \boldmath{$\mu$} \textbf{Sensitive Area} & \textbf{Instrumented Area} & \multirow{2}{*}{\textbf{Coverage}} \\
  & \textbf{(m)} & \textbf{(m}\boldmath{$^2$}\textbf{)} & \textbf{(m}\boldmath{$^2$}\textbf{)} & \\
\midrule
LHAASO & 4410 & 4.2~$\times$~10$^4$ & 10$^6$ & 4.4~$\times$~10$^{-2}$ \\
 TIBET AS$\gamma$ & 4300 & 4.5~$\times$~10$^3$ & 3.7~$\times$~10$^4$ & 1.2~$\times$~10$^{-1}$ \\
 KASCADE & 110 & 6~$\times$~10$^2$ & 4~$\times$~10$^4$ & 1.5~$\times$~10$^{-2}$ \\
 CASA-MIA & 1450 & 2.5~$\times$~10$^3$ & 2.3~$\times$~10$^5$ & 1.1~$\times$~10$^{-2}$ \\
\bottomrule
\end{tabular} 
\end{table}
%%%%%%%%%%%%%%%%%%%%%%%%%%%%%%%%%%%%%%%%%%%%%%%%%%%%%%%%%
%
Generally speaking, near the depth of the maximum of the shower development, the number of secondary charged particles is almost independent of the mass of the primary particle, and the shower fluctuations are at minimum. For the knee energy region, this depth corresponds to $\approx$5000 m asl.
Therefore, these extreme altitudes are suitable to have good energy resolution, to reconstruct the primary energy in a mass-independent way and to study  the shower core region in great detail, where the hadronic component feeds the e.m.\ one deep in the atmosphere. As demonstrated by the ARGO-YBJ \cite{bartoli2012} and Tibet AS$\gamma$ \cite{amenomori2006} experiments, observables related to the shower core properties are almost independent on the details of hadronic interaction models.
At high altitudes, due the low energy threshold ($\approx$TeV), it is possible to cross-check the fluxes with direct measurements on a wide energy range (ARGO-YBJ in the 5--250 TeV range). 
This cross-calibration is important due to the conflicting results obtained not only by ground-based detectors, but also by direct experiments.
In addition, the absolute energy scale can be calibrated at a level of 10\%, exploiting the so-called \emph{``Moon Shadow''} technique \cite{bartoli2011}.

On the other hand, experiments located deep in the atmosphere enhance the differences in the longitudinal development of EAS of different primary masses, as the shower is sampled well beyond its maximum. Therefore, the ratio N$_e$/N$_{\mu}$ is, in principle, more suitable for elemental composition studies.
However, shower fluctuations are much larger, making it difficult to interpret the data.
In addition, the reconstruction of the energy is typically strongly model-dependent because `a priori' assumptions on the primary composition are needed, and the calibration of the absolute energy scale is one of the major open issues.
\newpage
The great variety of layouts, observables, and reconstruction procedures to infer the elemental composition is at the origin, in part, of the conflicting results reported by different ground-based experiments.
Arrays focused on the investigation of the knee region operated so far are also characterized by a limited size of the instrumented area. They collected limited statistics above 10$^{16}$ eV, and were, therefore, unable to give a conclusive answer to the origin of the knee. The poor sensitivity to elemental composition, due to the small statistics, prevents  discrimination against different mass groups, and only general trends can be investigated in terms of the evolution of $\left<\ln A\right>$ or of \emph{``light''} and \emph{``heavy''} components with energy.

\section{Extensive Air Showers: The Heitler-Matthews Model}
\label{sec:easmodel}
%\medskip

A general idea of the main characteristics of EAS and of how different nuclei produce showers with different properties can be obtained from some relatively simple arguments, as suggested by Heitler \cite{heitler} and Matthews \cite{matthews}.
This toy model is useful to show how different observables depend on the primary mass and energy, and why certain techniques have historically been used to study elemental composition or to reconstruct the energy spectrum.
Nevertheless, detailed Monte Carlo simulations must be used to describe quantitatively all the characteristics of these random processes, with particular care to the role of shower fluctuations.

\begin{figure}[H]
\includegraphics[width=0.95\textwidth]{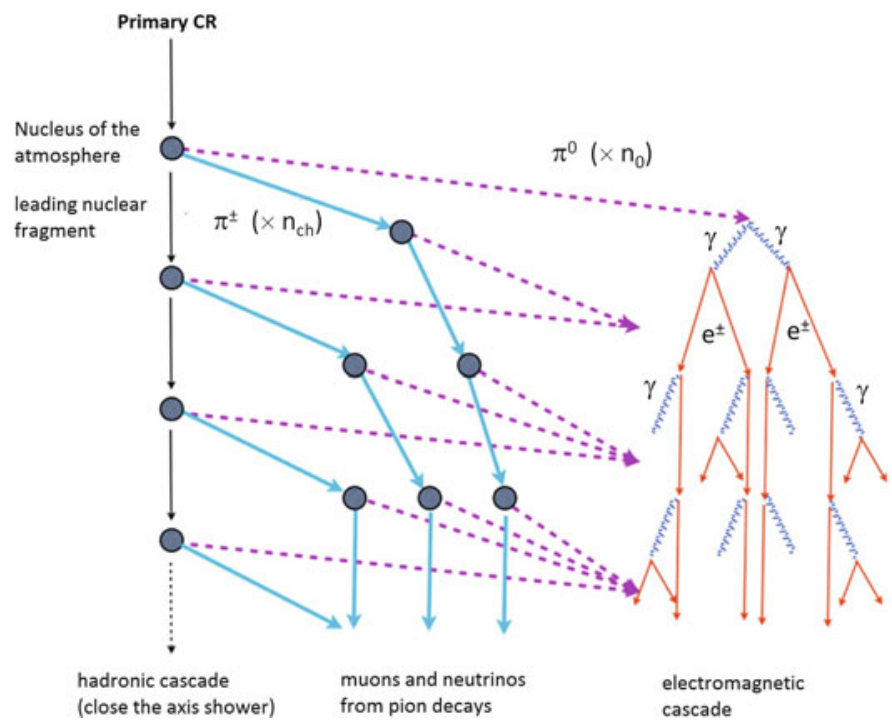}
\caption{Schematic evolution of cascades initiated by a CR particle. At each step, roughly 1/3 of the energy is transferred from the hadronic cascade to the e.m.\ one. Figure taken from \cite{letessier2011}.
    \label{fig:eas-schema}}
\end{figure}

In a nutshell, the collision of a primary CR with a nucleus of the atmosphere produces one large nuclear fragment and many charged and neutral pions (with a smaller number of kaons) (Figure~\ref{fig:eas-schema}) \cite{letessier2011}. A significant fraction of the total energy is carried away by a single \emph{``leading''} particle. This energy is unavailable immediately for new particle production. 
Roughly speaking, half of the energy of the primary particle is transferred to the nuclear fragment and the other half is taken by the pions (and kaons). The fraction of energy transferred to the new shower particles is referred as \emph{inelasticity}. Accurate description of the leading particles is crucial because these high-energy nucleons feed energy deeper into the atmosphere.
Approximately equal number of positive, negative and neutral pions are produced.
The e.m.\ component, the most intense of an EAS, is produced by the photons coming from the decay of the neutral pions.
At each interaction before the charged pions decay, nearly a third of the hadronic component energy is released into the e.m.\ one. 

As the number of particles increases, the energy per particle decreases. They will also scatter, losing energy, and many will range-out. 
Thus, the number of particles (or, with less ambiguities in the definition, the quantity of energy transferred to secondaries and eventually released into the atmosphere) will reach a maximum at some depth X$_{max}$ which is a function of energy, of the nature of the primary particle and of the details of the interactions of the particles in the cascade. 
After that, the energy/particle is so degraded (will be below some \emph{``critical energy''}) that energy losses dominate over particle multiplication process, and the shower ``size'' will decrease as a function of depth: it grows `old'. 
Once the pions have reached an energy which is low enough, they will decay into muons and neutrinos ($\pi^+\to\mu^+\nu_{\mu}$ or $\pi^-\to\mu^-\bar{\nu}_{\mu}$). The resulting muons propagate unimpeded to the ground. 
The muon cascade grows and maximizes, but the decay is slower as a consequence of the relative stability of the muon and small energy losses by ionization and pair production.

%%%%%%%%%%%%%%%%%%%%%
%%%%%%%%%%%%%%%%%%%%%

These are the most common processes, but not at all the only ones. As an example, successive hadronic interactions of the primary CR, interactions/decays of kaons and muon decays, multiple scattering and production angles must  also be considered (see, for example, the the Ref. \cite{gaisser}).
Only detailed simulations with Monte Carlo methods are able to describe all the characteristics of these random processes.

Historically, one of the main problems in analyzing data from shower arrays was related to the fact that each experiment used its own simulation of shower development and detectors. This made difficult the comparison of the results and the understanding of their differences.
Starting in the 1990s, all experiments began to use the same Monte Carlo simulation code CORSIKA \cite{corsika}, a framework containing different hadronic interaction models to describe the shower development in the atmosphere, and the software GEANT~\cite{geant} to simulate the detectors operated in the arrays.
 Over the years, other simulation codes have been developed, in particular to describe the development of showers at ultra-high energies, such as AIRES \cite{aires}. 
 The main characteristics of hadronic interactions that are relevant for EAS physics are: cross-sections (p--air, $\pi$--air, N--air), inelasticity of the collisions, multiplicity/composition of secondaries, transverse momentum distribution, fraction of diffractive dissociation. 

New data coming from the LHC (at an energy $E_{lab}$$\sim$$10^{15}$ eV) allowed to improve the models even if some points remain critical. 
In fact, the situation is much worse than it may appear from energy considerations.
Measurements at colliders are limited to an angular region that excludes the beam pipe (the so-called \emph{'central region'}), and therefore a very large majority of the high-energy particles that are emitted at small angles (in the so-called \emph{'forward region'}) are unobservable. In EAS physics the forward region is the most relevant because the high-energy particles feed energy in the shower down in the atmosphere.
Therefore, models tuned to accelerator measurement in the central region are extrapolated to describe the interactions of CRs.

Nevertheless, this simple toy model predicts the basic features of EAS development.
In the following, the e.m. and hadronic processes will be described separately in more~detail.

\subsection{Electromagnetic Showers}

\begin{figure}[H]
\includegraphics[width=0.9\textwidth]{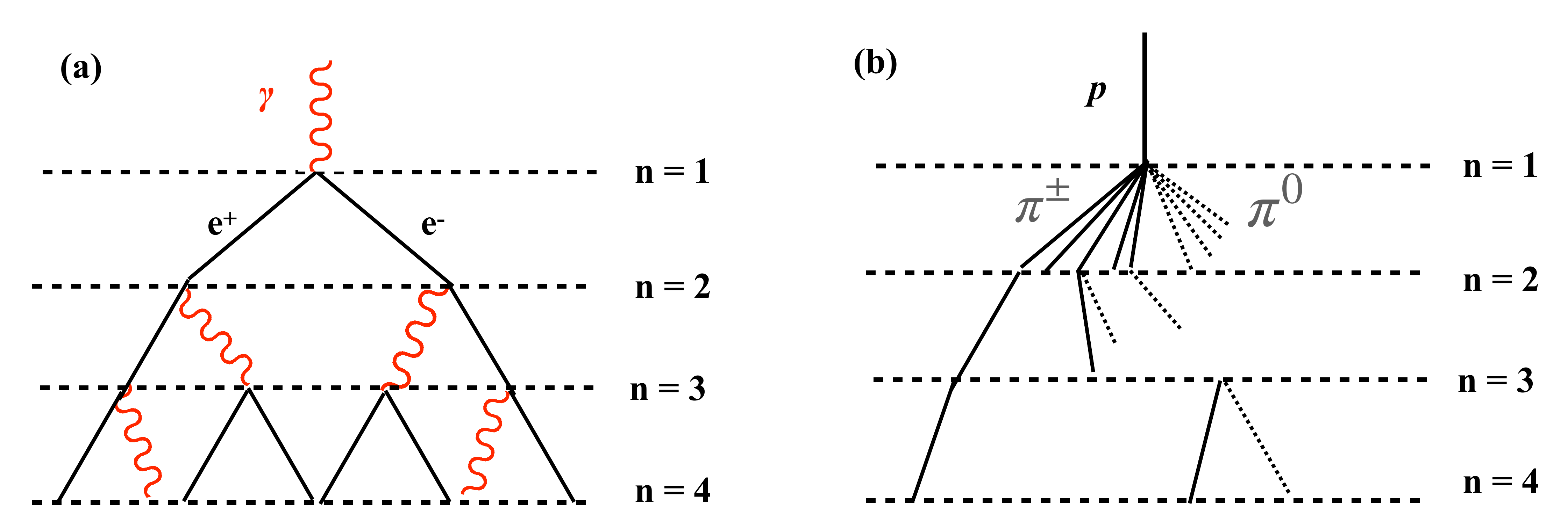}
\caption{Schematic view of an e.m.\ cascades (\textbf{a}), and of a hadronic shower (\textbf{b}). In the hadron shower, dashed lines show $\pi^0$ which do not re-interact but decay, producing e.m.\ sub-showers.}
\label{fig:eastoymodel}       % Give a unique label
\end{figure}

The main features of an e.m.\ shower profiles can be described within the simple Heitler's toy model of particle cascades \cite{heitler}.
Let us suppose that a particle (electron, positron or photon) with energy E$_0$ splits its energy equally into two particles after traveling a radiation length X$_0$ in air, and let this process be repeated by the secondaries (see Figure~\ref{fig:eastoymodel}).
%
%%%%%%%%%%%%%%%%%%%%%%%%%%%%%%%%%%%%
%%%%%%%%%%%%%%%%%%%%%%%%%%%%%%%%%%%%
%
Let $X$ describe the depth in the atmosphere and define the depth at which the average CR starts interactions with the atmosphere to be $X$ = 0 g/cm$^2$.
After $n$ radiation lengths, we obtain a particle cascade which has evolved into $N$ = 2$^n$ particles of equal energy $E$ = $E_0/N$.
Multiplication stops when the energies of the particles are too low for pair production or bremssthralung. This energy is the critical energy $\varepsilon_c^{em}$ in the air ($\approx$80 MeV, below which the collisional energy losses are dominant).

The maximum number of particles $N_{max}$ is reached at this moment, when all particles have the same energy $\varepsilon_c^{em}$, $E_0$ = $\varepsilon_c^{em} \cdot N_{max}$.
The depth $X_{max}$ at which the shower reaches the maximum size is  $X_{max}$ = $n_{max} \cdot X_0$, where $n_{max}$ is the number of radiation lengths required for the primary energy to be reduced to $\varepsilon_c^{em}$.

Since $N_{max}$ = $2^{n_{max}}$, we have
\begin{equation}
n_{max} = \ln\bigg(\frac{E_0}{\varepsilon_c^{em}}\bigg) \cdot \frac {1}{\ln 2}
\end{equation}
so that
\begin{equation}
\label{eq:xmax}
X^{em}_{max} = \frac{X_0}{\ln 2} \cdot \ln \bigg(\frac{E_0}{\varepsilon_c^{em}}\bigg).
\end{equation}

Finally, it is interesting to estimate the \emph{elongation rate} $\Lambda$, that is, the rate of increase
of $X_{max}$ with the primary energy. From the relation (\ref{eq:xmax}), we have
\begin{equation}
\Lambda^{em}=\frac{d\, X_{max}}{d\, \log_{10} E_0} = 2.3\cdot X_0 = 85\>\> \text{g/cm$^2$}\hspace{0.5cm} \text{per decade of energy}.
\end{equation}

This simple model predicts two basic features of e.m.\ shower development:
\begin{itemize}
\item $N_{max}$ increases proportional to the primary energy \emph{E}$_0$, \emph{N}$_{max}$ = $\frac{E_0}{\varepsilon_c^{em}}$.
\item $X_{max}$ increases logarithmically with primary energy, at a rate of 85 g/cm$^2$ per decade of energy.
\end{itemize}

\subsection{Hadronic Showers}
\label{sect:hadint}

Air showers initiated by protons have been modeled by different authors (see, for example, the the Ref. \cite{horandel2007,kampert2012,mollerach2018}) following the Matthews approach \cite{matthews}, similar to the Heitler one. The main differences with the e.m.\ cascades are
\begin{enumerate}
\item in the hadronic interactions a large number of secondary particles are produced. At high energy, the total multiplicity of particles per collision, $N_{\rm tot}$, reaches values of several tens with the consequence that although the hadronic interaction lengths are larger than the e.m.\ radiation one (at PeV energies $\lambda_{p\,{\rm -air}}\simeq 80$~g~cm$^{-2}$), the showers develop faster than in the e.m.\ case;
\item in a hadronic interaction only a fraction of the energy E$_0$ is available for secondary particle production. A single leading particle, the highest energy secondary produced in the interaction, carries a fraction $(1-\kappa)E_0$ deep in the atmosphere, where $\kappa$ is the so-called \emph{inelasticity}. Therefore, a fraction $\frac{2}{3}\kappa E_0$ is used to produce $N_{ch}$ charged pions, and a fraction $\frac{1}{3}\kappa E_0$ goes via neutral pions into the e.m.\ component;
\item the critical energy $\varepsilon_c^\pi$ is defined as the energy at which the decay and the hadronic interaction probabilities are equal and further particle production by $\pi^\pm$ ceases. 
\end{enumerate}

According to the Ref. \cite{matthews}, constant values $N_{ch}$ = 10, corresponding to an energy of about 100~GeV, and $\varepsilon_c^\pi$ = 20 GeV are adopted in the following.

Protons travel one interaction length and interact producing $N_{tot}$ pions, all having equal energies, $N_{ch}$ are charged and $\frac{N_{tot}}{3}=\frac{1}{2}\cdot N_{ch}$ neutral, which immediately decay into photons, initiating e.m.\ showers. As for the e.m.\ cascade, we assume equal division of energy during particle production.

In turn, the charged pions can decay in muons and neutrinos and hence, as long as their decay length remains larger than their interaction length, they will re-interact rather than decay.  This happens for $\gamma c\tau_{\pi}>\lambda_{\pi\,{\rm -air}}/\rho_{\rm air}$, with the Lorentz factor $\gamma=E_{\pi}/m_\pi$, the charged pion lifetime $\tau_\pi\simeq$ 26 ns and $\lambda_{\pi\,{\rm -air}}\simeq 1.5\lambda_{p\,{\rm -air}}\simeq$ 120 g cm$^{-2}$ (since the $\pi p$ cross-section is about 2/3 the $pp$ cross-section). 
This implies that pions will re-interact as long as their energy satisfies \emph{E} $>$ \emph{E}$_{\rm d}\simeq$ 100~GeV(10$^{-4}$ g cm$^{-3}/\rho_{\rm air})$. Hence, at the  heights above 10 km, where the initial development of the shower takes place, $\pi^\pm$ will re-interact for energies greater than $\sim$20--30 GeV \cite{mollerach2018}.  

After $n$ interactions, the $N_{\pi}$ = $(N_{ch})^n$ charged pions produced carry a total energy of $(\frac{2}{3})^n \cdot E_0$. The energy per charged pion after $n$ interactions is then $E_{\pi^\pm}$ = $\frac{E_0}{(3/2 N_{ch})^n}$.
The remainder of the primary energy goes into the e.m.\ component from $\pi^0$ decays
\begin{equation} 
E_{\rm em}\simeq E_0\left[1-\left(\frac{2}{3}\right)^n\right].
\end{equation}

After only six interactions, about 90\% of the initial energy is transferred to the e.m.\ component of the shower, with the remaining 10\% being essentially the muons and neutrinos from the charged pion decays.
As a consequence, most of the energy of an air shower can be observed in its e.m.\ component.
This is the so-called \emph{calorimetric energy} which allows to estimate the primary energy with good accuracy to detectors able to observe the longitudinal air shower development.

Assuming that at $\varepsilon_c^\pi$, all pions decay, the number of muons is $N_{\mu}$ = $N_{\pi^\pm}$ = $(N_{ch})^{n_{c}}$, where $n_{c}$ is the number of interaction lengths required for the charged pion's interaction length to exceed its decay length 
\begin{equation} \label{nceq}
 n_c= \frac{\ln (E_0/\varepsilon_c^\pi)}{\ln(\frac{3}{2}N_{ch})}
    = 0.85\,\lg\left(\frac{E_0}{\varepsilon_c^\pi}\right).
\end{equation}

Thus, the total energy is divided into two channels, hadronic and electromagnetic
\begin{equation} \label{enecons}
E_0 = E_{em} + E_h = \varepsilon_c^{em}\cdot N_e + \varepsilon_c^{\pi}\cdot N_{\mu}.
\end{equation}

This equation represents \emph{energy conservation}, apart from a fraction of a few percent of the primary energy spent in the neutrino component. The relative magnitude of the contribution from $N$$_{\mu}$ and $N$$_e$ does not depend on the details of the model, but only on the respective critical energies, the energy scales at which e.m.\ and hadronic multiplication ceases.
An important conclusion of this description of the hadronic cascades is that the energy is given by a linear combination of muon and electron sizes. This result is insensitive to fluctuations in the division of energy between the hadronic and e.m.\ channels and independent on the mass of the primary particle.

The number of muons is given by
\begin{equation}
\ln N_{\mu}=\ln N_{\pi^\pm} = n_c \ln N_{ch} = \frac{\ln (E_0/\varepsilon_c^{\pi})}{\ln (3/2 N_{ch})}\cdot \ln \big(N_{ch}\big)
= \beta \cdot \ln \bigg(\frac{E_0}{\varepsilon_c^{\pi}}\bigg).
\end{equation}

Following \cite{matthews}, we can estimate $\beta=\frac{\ln\, (N_{ch})}{\ln\, (3/2 N_{ch})}=0.85$ for $E_0$ in the range 10$^{14}$--10$^{17}$ eV, obtaining
\begin{equation} \label{nmuprot}
N_{\mu} = \bigg( \frac{E_0}{\varepsilon_c^{\pi}}\bigg)^{\beta}=\bigg( \frac{E_0}{\varepsilon_c^{\pi}}\bigg)^{0.85} \sim 9900 \bigg(\frac{E_0}{10^{15} \text{eV}}\bigg)^{0.85}.
\end{equation}

Including inelasticity in the Heitler model \cite{matthews} changes the parameter $\beta$
\begin{equation} \label{betakappa}
 \beta=\frac{\ln\, (N_{ch})}{\ln\, (3/2 N_{ch})} \to \frac{\ln[1+N_{ch}]}{\ln\left[(1+N_{ch})/(1-\frac{1}{3}\kappa)\right]}
      \approx1-\frac{\kappa}{3\ln(N_{ch})}=1-0.14\kappa.
\end{equation}

The elasticity for the most energetic meson in pion--air interactions yields $(1-\kappa)$ between 0.26 and 0.32, resulting in $\beta=0.90$.

The electronic size can be calculated by inserting the expression (\ref{nmuprot}) for the muon size in the energy conservation relation (\ref{enecons}) 
\begin{equation} \label{emfracteq}
 \frac{E_{em}}{E_0}= \frac{E_0-N_\mu \varepsilon_c^\pi}{E_0}
                   = 1-\left(\frac{E_0}{\varepsilon_c^\pi}\right)^{\beta-1}.
\end{equation}

The e.m.\ fraction is 66\% at $E_0=10^{15}$~eV, increasing to 83\% at $10^{18}$~eV for proton-induced showers. 

Therefore, the number of electrons at a maximum shower for proton-induced \mbox{showers~is} 
\begin{equation} \label{neeq}
   N_e = \frac{E_{em}}{\varepsilon_c^{em}} = \frac{E_0}{\varepsilon_c^{em}} - \frac{\varepsilon_c^{\pi}}{\varepsilon_c^{em}}\bigg(\frac{E_0}{\varepsilon_c^{\pi}}\bigg)^{\beta} 
   \approx \frac{E_0}{\varepsilon_c^{em}} = N_{e|_{max}}^p.      
\end{equation}

The approximation is justified at high energies when the fraction of energy transferred to muons is small \cite{kampert2012}.

In the framework of the \emph{superposition model}, each nucleus is taken to be equal to $A$ individual single nucleons, each with energy $E_0/A$ and each acting independently.
The shower resulting from the interaction of the primary nucleus $A$ can be treated as the sum of $A$ proton-induced independent showers all starting at the same point.
Thus, while a proton creates one shower with energy $E_0$, an iron nucleus of the same total energy is expected to create the equivalent of 56 proton showers, each with reduced energy ($E_0/56$). The average properties of showers are well reproduced by this model, though the fluctuations are clearly underestimated and can be studied only with detailed Monte Carlo simulations of the intra-nuclear cascade.
The superposition of \emph{A} independent showers naturally explains why the shower-to-shower fluctuations are smaller for shower initiated by nuclei as compared to proton showers.

By substituting  the lower primary energy $({E_0}/A)$ into the previous expressions and summing $A$ such showers, we obtain the following relations for the number of electrons and muons in a shower induced by a nucleus $A$:
\begin{equation} \label{neeq} 
  N_{e|_{max}}^A = A \left(\frac{E_0/A}{\varepsilon_c^{em}}\right)  = N_{e|_{max}}^p      
\end{equation} 
\begin{equation} \label{nmueq}
 N_{\mu |_{max}}^A = \left(\frac{E_0}{\varepsilon_c^\pi}\right)^\beta A^{1-\beta}=N_{\mu |_{max}}^p A^{1-\beta} 
         \approx1.69\cdot10^4 \cdot
	 A^{0.10} \left(\frac{E_0}{1~\mbox{PeV}}\right)^{0.90}.
\end{equation}

From these relations valid \emph{at shower maximum} follows:
\begin{enumerate}
\item The number of electrons is equal for all primary masses A, that is, is independent of the composition. Therefore, \emph{the shower size $N_{e|_{max}}^A$ can be used as an estimator of the energy};
\item The number of muons $N_{\mu |_{max}}^A$ increases with the mass of the primary particle with $A^{1-\beta}$$\sim$$A^{0.1}$. Accordingly, iron-induced showers contain about 1.5 times as many muons as proton showers with the same energy.  
In fact, in a shower induced by a nucleus A, due to the smaller energy per nucleon (\emph{E}$_0$/\emph{A}), the secondary pions are less energetic. This favours a pion decay as well as an interaction of heavier nuclei higher in atmosphere, where the air density is smaller. \emph{The number of muons can be used to infer the mass of the primary particle}. 
Moreover, the evolution of the muon number with energy, d\emph{N}$_{\mu}$/d ln\emph{E}, is a good tracer of changes in the primary composition. In fact, a constant composition gives d\emph{N}$_{\mu}$/d ln\emph{E} = $\beta$ and any departure from that behavior can be interpreted as a change of the average mass of the primaries, in a similar way as with the elongation rate of the longitudinal development.
\item The muon size grows with primary energy more slowly than proportionally, $\beta$$\sim$$0.90$.
\end{enumerate}

A large number of ground-based arrays studying the knee energy region are located deep in the atmosphere and do not sample the number of electrons at shower maximum. Therefore, the experimental situation is not ideal because the size, used to recover the energy of the primary particle, is mass-dependent, as discussed in  Section~\ref{enemassrec}. Only experiments located at extreme altitude (above 4000 m asl) observe the electrons in the shower maximum region for near-vertical showers with an energy in the PeV range.

Deeper in the atmosphere, arrays measure only the attenuated size
\begin{equation}
N_{e|_{ground}}\approx N_{e|_{max}}\cdot \exp \left(-\frac{\Delta X}{\Lambda}\right)
\end{equation}
where $\Delta X$ is the distance of the shower maximum from the ground and $\Lambda$ $\approx$ 60 g/cm$^2$ is the attenuation length of the electron size after the shower maximum. 
Since heavy nuclei reach the maximum of longitudinal development at smaller depths than light ones, on the ground we have a larger electron number for air showers initiated by light particles.
This implies that, due to the steeply falling CR spectrum, showers of equal $\ln N_e$ are enriched in light elements.

\subsection{Longitudinal Development}

The longitudinal development of a hadronic shower is dominated by the parallel e.m.\ sub-showers produced by the $\pi^0$ decays in the first interaction, at an atmospheric depth $X^* = \lambda_{p-air} \cdot \ln\, 2\approx$ 55 g/cm$^2$. In a good approximation following cascades can be neglected. The energy of the single photon is \emph{E}$_{\gamma}$ = $\frac{E_{\pi^0}}{2}$ = $\frac{E_0}{3}\frac{2}{N_{ch}}  = \frac{E_0}{3 N_{ch}}$.

From Equation (\ref{eq:xmax}), we have
\begin{align*}
X^p_{max} & = X^* + X_0 \cdot \ln \bigg(\frac{E_0}{3 N_{ch}\cdot \epsilon_c^{em}}\bigg) \\
 & = X^* + X_{max}^{em} - X_0 \cdot \ln (3\, N_{ch})\>\>\>\>\>\>{\rm g/cm^2}
\end{align*}
where $X_{max}^{em}$ is the atmospheric depth of the maximum of $\gamma$-induced showers with \emph{E}$_0$ primary energy and \emph{N}$_{ch}$ is the multiplicity of charged pions in the first interaction.
The elongation rate for showers induced by protons is then 
\begin{equation}
\Lambda^p = \Lambda^{\gamma} + \frac{d}{d\, \log_{10} E_0}\bigg[ X^* - X_0\cdot \ln (3\, N_{ch})\bigg] = 58 \>\>\>\>\>\>{\textrm{g/cm$^2$ per decade,}}
\end{equation}
reduced from the elongation rate for purely e.m.\ showers. This estimation verifies Linsley's elongation rate theorem \cite{linsley77}, which points out that e.m.\ showers represent an upper limit to the elongation rate of the hadronic showers.
The shower maximum is expected to be influenced by the elasticity of the first interaction, ($1-\kappa$) = \emph{E}$_{lead}$/\emph{E}$_0$, where \emph{E}$_{lead}$ is the energy of the leading particle. 
For interactions with ($1-\kappa$) $>$ 0.5 most of the primary energy will be transferred deeper into the atmosphere and correspondingly the shower maximum will be~deeper.%please confirm if it should be minus sign

\textls[-15]{The extrapolation to a primary particle with mass \emph{A} with the superposition model~yields}
\begin{equation}\label{xmaxa}
X_{max}^A = X_{max}^p - X_0\cdot \ln A
\end{equation}

Detectors able to observe the longitudinal air shower development can estimate the primary energy with good accuracy measuring the so-called \emph{calorimetric energy}, that is, the energy of the e.m.\ component.
With this estimator of the energy of the primary particle, the orthogonal variable sensitive to its primary mass is the depth of the shower maximum in terms of the number of particles, \emph{X}$_{max}$.

Therefore:
\begin{itemize}
\item $X_{max}$ is smaller for heavier nuclei (logarithmic dependence on $A$)
\item $X_{max}$ is the same for same $E_0/A$ but different $E_0$. As a consequence, the proton-induced showers result, on average, in a larger number of particles at the observation level compared to iron-induced events. However, the shower-to-shower fluctuations are as large as the shift of $X_{max}$ between proton and iron thus limiting an event-by-event assignement of a primary mass.
\end{itemize}

Despite the simplicity and the approximations of the toy model, the main characteristics of the EAS development are quite well reproduced. Obviously, a detailed description of the cascade, in particular for what concern the role of fluctuations, can be provided only by detailed Monte Carlo simulations.

\subsection{Energy and Mass}
\label{sect:crmass}

The relevance of muon measurements to the question of the primary composition has been first remarked by the Institute for Nuclear Studies (INS) group in Tokyo \cite{fukui1960}. 
They were the first group to point out the key information that the mass of the primary particle could be derived from a study of plots of muon versus electron number.

Due to the intuitive relation between shower to shower fluctuations and primary mass, the study of fluctuations in the muon number distributions was historically the first method employed to study the primary CR mass composition \cite{fukui1960, matano1963,khristiansen1963}.
The narrowing of the distribution of \emph{N}$_{\mu}/\emph{N}_{e}$ was considered to be due partly to the change in the composition of primary particles with energy \cite{matano1963}.

On general grounds, the elemental composition can be investigated if the total size and the muon component depend differently from the primary energy.
If we assume that their dependences from the energy of a primary proton \emph{E}$_0$ can be described as
\begin{equation}
N_e\propto E_0^{\beta_e}, \hspace{1.5cm} N_{\mu}\propto E_0^{\beta_{\mu}},
\end{equation}
for a nucleus $A$, we have
\begin{equation}
N_e\propto \bigg(\frac{E_0}{A}\bigg)^{\beta_e}\cdot A, \hspace{1.5cm} N_{\mu}\propto\bigg(\frac{E_0}{A}\bigg)^{\beta_{\mu}}\cdot A,
\end{equation}
with a mass-number dependency of the type $1-\beta_e$ and $1-\beta_{\mu}$, respectively.
The relation $N_{\mu}/N_e$ can be easily deduced
\begin{equation}
N_{\mu}\propto N_e^{\beta_{\mu}/\beta_e} A^{1-(\beta_{\mu}/\beta_e)}.
\end{equation}

In 1962, Linsley, Scarsi, and Rossi working at the MIT Volcano Ranch Station observed, for the first time, a muon/electron correlation: $N_{\mu}$$\sim$$A^{1-\alpha}\cdot (N_e)^{\alpha}$, thus establishing that the muon size is a mass-sensitive observable \cite{linsley1962}.

The Equation (\ref{neeq}) can be transformed to obtain the energy $E_0$ to be introduced in the relation (\ref{nmueq}) to obtain $\beta_{\mu}/\beta_e$$\sim$$0.86$. The muon size for a given mass $A$ as a function of the total size $N_e$ is then
\begin{equation}
N_{\mu}\propto N_e^{0.86} A^{0.14}.
\end{equation}

In a similar way, we can obtain the muon size as a function of the total size for a given primary energy $E_0$.
The Equation (\ref{neeq}) is transformed to obtain the mass $A$ which in turn is introduced in the relation (\ref{nmueq}) 
\begin{equation}
N_{\mu}\propto \bigg(\frac{E_0}{1\>\>\text{PeV}}\bigg)^{3.17} N_e^{-2.17}.
\end{equation}

In experiments with ground-based arrays the reconstructed number of muons and electrons are plotted in a $\ln N_{\mu} - \ln N_{e}$ plane to recover the energy and mass of the primary particle.
This diagram, when combined with detailed shower simulations, proved to be a powerful tool for extracting information on primary mass.

Therefore, it is interesting to study the electron-to-muon ratio at shower maximum
\begin{equation}\label{nenmuratio}
\frac{N_e}{N_{\mu}}\approx 35.1\cdot \bigg(\frac{E_0}{A}\bigg)^{0.15}.
\end{equation}
with the energy in PeV.
This ratio depends on the energy per nucleon $E$$_0$/$A$ of the primary particle, thus showing that $N$$_e$/$N$$_{\mu}$ can be used to infer the mass of the primary particle if the energy is measured with a different, independent observable.

We can use the relation (\ref{nenmuratio}) to investigate the sensitivity of EAS arrays to the primary mass A \cite{horandel2007,horandel2008}
\begin{equation}
\lg\bigg(\frac{N_e}{N_{\mu}}\bigg)= 1.54 + 0.15\cdot\lg\bigg(\frac{E_0}{1\>\>\text{PeV}}\bigg) - 0.065\cdot \ln A = C - 0.065\cdot \ln A
\end{equation}

If the energy is reconstructed from another independent observable, the mass of the primary CR can be determined by measuring the ratio $N$$_e$/$N$$_{\mu}$.
Therefore, the relative error on the electron-to-muon ratio is 
\begin{equation}
\frac{\Delta(N_e/N_{\mu})}{N_e/N_{\mu}}\sim 0.15\bigg[\frac{\Delta E_0}{E_0}+\frac{\Delta A}{A}\bigg]\sim 0.15\bigg[\frac{\Delta A}{A}\bigg]
\end{equation}
with the consequence that to measure the elemental composition with a resolution of one unit in $\ln A$ the relative error on $N_e/N_{\mu}$ must be $\approx$15\%.
A resolution of one unit in $\ln A$ in principle allows to reconstruct 4 (or 5 ?) different mass groups: p, He, CNO, MgSi (?) and Fe. The large shower-to-shower fluctuations often only allow one to trace the light and heavy components or the parameter $\left<\ln A\right>$ with energy.
Similarly, from the relation (\ref{xmaxa}) follows that the position of the shower maximum must be determined with a resolution of about one radiation length X$_0$$\sim$37 g/cm$^2$ to have a resolution in $\ln A$ of one unit.

\section{Reconstruction of the Energy and Mass of the Primary Particle}
\label{enemassrec}

The crucial point in air shower observations with EAS arrays is the reconstruction of the primary particle properties (especially energy and mass number) from the measured quantities.
In fact, analysis of shower data consists in the disentanglement of a threefold problem involving primary energy, primary mass and modelling of hadronic interactions (for a discussion about hadronic interactions in CR physics see, as an example, refs. \cite{lipari2014,riehn2020} and references therein).
An intrinsic ambiguity affects the interpretation of data. 
Different combinations of the two following elements can produce similar showers. 
As an example, a \emph{``short''} shower can be produced by a large cross-section, high inelasticity or heavy primary mass.
On the contrary, a \emph{``long''} shower, penetrating deeper in the atmosphere, can be produced by small cross-section, low inelasticity or light mass.
\begin{itemize}
\item[(1)] \textls[-20]{shower development, mainly governed by the inelasticity and by the inelastic cross~section}
\item[(2)] elemental composition of the primary flux, that we don't know and want to measure
\end{itemize}

Strictly speaking, when operating with shower arrays there are no observables directly related to the mass of the primary particle, and its measurement is very indirect. 
We note, however, that the Cherenkov light emitted by a primary heavy nucleus high up in the atmosphere (the so-called \emph{``direct Cherenkov light''}) is directly related to the charge (and therefore to the mass) of the primary particle \cite{kieda2001,hess-fe2007}. Since this light is proportional to $Z^2$, heavy nuclei are more suited for detection.
Charge resolution is about 10\% for Z $>$ 10. The main limitation is that it can only be used over a small energy range for each atomic charge~Z. 

The majority of experiments with shower arrays can therefore apply only `statistical' methods according to a classical scheme:
\begin{enumerate}
\item From the experimental data, via some phenomenological functions determined by Monte Carlo simulations for the particular array, the measured observables (\emph{N}$_e$, \emph{N}$_{hadr}$, \emph{N}$_{\mu}$, \emph{X}$_{max}$, $\ldots$) are reconstructed. 
\item The distributions of such quantities are compared with those extracted from a detailed simulation of the EAS development in the atmosphere in which a trial CR spectrum is~used. 
\item The input spectrum is varied in order to optimize the agreement between the reconstructed and calculated distributions of measured observables.
\end{enumerate}

Therefore, a typical data analysis consists in finding a combination of primary energy spectrum, elemental composition and hadronic interaction characteristics to obtain a consistent description of the experimental results. 
Clearly, this is not a measurement, but only a consistency check of some trial models.
In case of discrepancy, it is difficult to identify the origin; in case of agreement, is the parameter combination unique?

Due to the reduced resolution in the measurement of the primary mass (see Section \ref{sec:easmodel}), the majority of shower arrays displayed the results only as a function of the total energy per particle with the so-called \emph{``all-particle''} energy spectrum, that is, as a function of the total energy per nucleus, and not per nucleon.
Any tentative to infere informations about elemental composition are limited, at most, to study the evolution of the \emph{``light''} (``proton-like'') or \emph{``heavy''} (``iron-like'') components as a function of the energy, with results which critically depend on Monte Carlo predictions.

In the last two decades, a number of multi-component experiments have started to measure, with high statistics, at the same time, different shower observables, on an event-by-event basis. 
This fact allowed to exploit sophisticated analysis techniques to infer the characteristics of the primary particle by measuring the correlation between different components (for a review see, for example, the Refs.~\cite{haungs2003,kampert2012} and references therein).

In a nutshell,
\begin{itemize}
\item \emph{How to obtain the energy spectrum in shower arrays?}

This is the first step in the analysis of CR data.
We measure the spectrum in one observable and make a conversion to the energy spectrum. 
The observable typically used is the shower size because, as discussed in Section~\ref{sect:hadint}, the number of electrons at shower maximum is nearly independent on the primary mass: \emph{N}$_{e|_{max}}^A\approx$ \emph{N}$_{e|_{max}}^p$. However, surface detectors are usually  located deep in the atmosphere and do not measure the number of electrons at shower maximum. Beyond the maximum, the number of electrons is a mass-sensitive parameter, with a larger electron number for air showers initiated by light primaries, according to a relation of the type
\begin{equation}
N_e(E,A) = \alpha(A)\cdot E^{\beta(A)}
\end{equation}
where the parameters $\alpha$ and $\beta$ depend on the primary mass $A$. This implies a degeneracy in the reconstruction procedure because to recover the primary energy from the size spectrum we must assume a given elemental composition to be measured. 
If the composition changes in the investigated energy range, the relationship between the measured electron size and inferred energy will also vary.
The number of electrons in the core region has been used in some experiment, as well as the particle density at a suitable given distance from the shower axis, in some large arrays (see, for example, the Refs. \cite{disciascio-rev,cao2021nat}). In both cases this densities, according to Monte Carlo simulations, are nearly independent of the primary mass.

\item \emph{How do we measure elemental composition at ground?}\\
The inelastic cross-section $\sigma^{Fe-Air}_{inel}$ of iron at 1 PeV is about six times larger than for protons of equal energy. Hence, nuclei develop showers higher in atmosphere (smaller \emph{X}$_{max}$) than protons, dissipating their energy much faster.
Due to the shorter interaction length and the smaller energy per nucleon and because of the reduced attenuation of the muon component, nuclei-induced showers contain less particles in the e.m.\ component deep in the atmosphere, but they carry more muons than a proton shower of the same energy.
This is the basis of the electron-muon correlation method.
Therefore, as discussed in Section~\ref{sect:crmass}, the measurement of electron and muon contents simultaneously (with their fluctuations) has become the first and most commonly employed technique to infer the CR elemental composition with arrays. 
However, intrinsic shower to shower fluctuations limit mass resolution to a few mass groups (see Section~\ref{sect:crmass}) and electron and muon numbers are not independent. In addition, the muon component is heavy dependent on the details of the hadronic interactions and the results strongly depend on the particular model used to interpret the data. 

The other common technique, below 10$^{18}$ eV, involves the observation of the Cherenkov light and the study of its shape. In fact, the characteristics of the photon distribution depend on the depth of the shower maximum, therefore on the mass of the primary particle. The overall Cherenkov intensity provides a calorimetric measurement of the CR energy.
Cherenkov light has been measured, for instance, in hybrid experiments by ARGO-YBJ and Tunka apparatus.

The KASCADE multi-component array was the first experiment that claimed the measurement of the energy spectra of 5 different mass groups (p, He, CNO, MgSi, Fe) through a complex unfolding of the \emph{N}$_e$/\emph{N}$_\mu$ diagram \cite{kascade1,kascade2,kascade3}.
In the last two decades, other multi-component experiments measured a number of observables that, in principle, are mass-sensitive: steepness of the lateral distribution, characteristics of shower core region, distribution of the relative arrival times and angles of incidence of the muon component, characteristics of the lateral distribution of high energy muons (the so-called \emph{``muon bundles''}) measured underground, pulse shape and lateral distribution of the air Cherenkov light, depth X$_{max}$ of the shower maximum (see, for example, \mbox{ref. \cite{grieder,spurio}}).
But the study of the muon component has remained the most used technique.
\end{itemize}

\section{Elemental Composition in the 10\boldmath{$^{14}$} to 10\boldmath{$^{18}$} eV Region}
\label{sec:composition}

Several experimental results associate the knee with the bending of the light component (p and He), and are compatible with a rigidity-dependent cut-off \cite{aglietta2004,kascade1,kascade2,kascade3,garyaka2007,tanaka2012}. However, the flux of the different components vary significantly depending on the interaction model used to interpret the data \cite{kascade1,kascade2,kascade3}.
On the contrary, other results (in particular those obtained by arrays located at high altitudes) seem to indicate that  the knee of the all-particle energy spectrum is due to heavier nuclei and that the light component cuts off well below \mbox{1 PeV  \cite{disciascio-rev,argo-hybrid2,amenomori2006,casamia,basje-mas,aglietta2004}}. 

In this section the measurements of the light component energy spectrum in the 10$^{14}$ to 10$^{18}$ eV region will be presented by using different plots to point out the conflicting results between the experiments. 

\begin{figure}[H]
%\begin{center}
\includegraphics[width=0.99\textwidth]{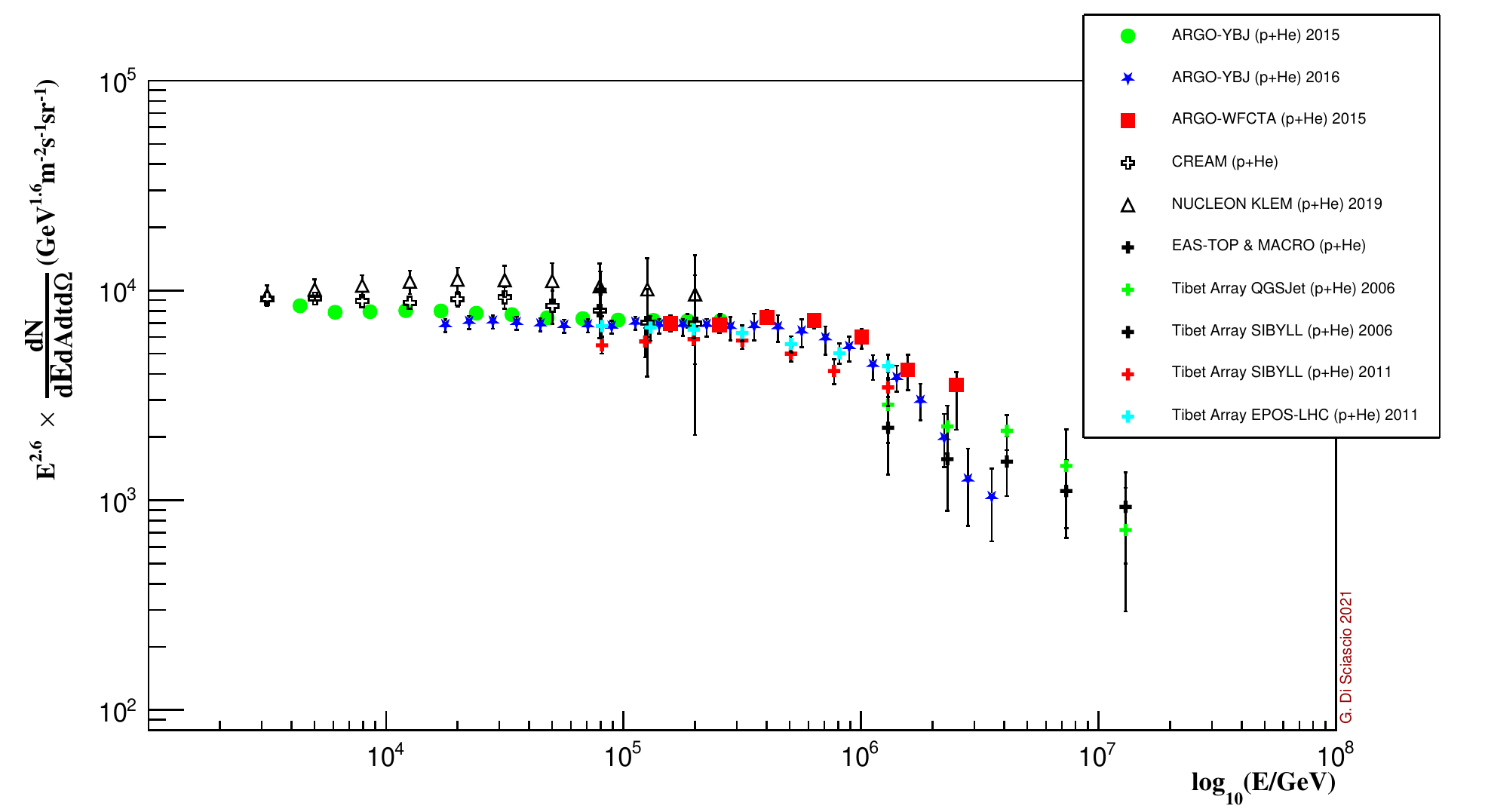}
\caption{Energy spectra of the light (p+He) component as measured by Tibet AS$\gamma$ \cite{amenomori2006,amenomori2011} and ARGO–YBJ \cite{disciascio-rev} experiments with different techniques and analyses, compared with results obtained in direct observations by CREAM \cite{cream2011} and NUCLEON \cite{nucleon2019}. 
    \label{fig:pheknee}}%MDPI: please   change the hyphen to minus sign, -1 should be −1 
   % \end{center}
\end{figure}

In Figure~\ref{fig:pheknee} the energy spectra of the light component as measured by Tibet AS$\gamma$ \cite{amenomori2006,amenomori2011} and ARGO-YBJ \cite{disciascio-rev} are shown. Both experiments are located in the YangBaJing Cosmic Ray Laboratory in Tibet (China) at 4300 m a.s.l. and did not exploit a measurement of the muon component to determine the elemental composition of the primary CR flux.
%
%%%%%%%%%%%%%%%%%%%%%
%%%%%%%%%%%%%%%%%%%%%

The Tibet AS$\gamma$ Collaboration reconstructed the energy spectrum studying the shower core region with a burst detector as well as with emulsion chambers.
The ARGO-YBJ experiment measured the CR energy spectra exploiting completely different and independent approaches \cite{disciascio-rev}:
\begin{itemize}
\item \emph{`Digital-Bayes' analysis}, based on the strip multiplicity, that is, the picture of the EAS provided by the RPC strip/pad system, in the few TeV--300 TeV energy range. The selection of light elements is based on the characteristics of the charged particle lateral distribution \cite{bartoli2012,bartoli15}.
\item \emph{`Analog-Bayes' analysis}, based on the RPC charge readout \cite{argo-bigpad}, covers the 30 TeV--10~PeV range. The energy is reconstructed (as in the previous analysis), on a statistical basis, by using a bayesian approach.
%\cite{argo-rm3knee}.
\item \emph{`Hybrid measurement'}, carried out by ARGO-YBJ and a wide field of view Cherenkov telescope, a prototype of the LHAASO telescopes, in the 100 TeV--3 PeV region. The selection of (p+He)-originated showers is based on two observables, the shape of the Cherenkov image and the particle density in the core region measured by the ARGO-YBJ central carpet. The energy is reconstructed by the telescope with a resolution better than 20\% \cite{argo-hybrid1,argo-hybrid2}.
\end{itemize}

All the results are in excellent agreement.
In the ARGO-YBJ experiment the selection of (p+He)-originated showers is performed not by means of an unfolding procedure after the measurement of electronic and muonic sizes, but on an event-by-event basis exploiting showers topology, that is, the lateral distribution of charged secondary particles. This approach is made possible by the full coverage of the central carpet, the high segmentation of the read-out and the high altitude location of the experiment that retains the characteristics of showers lateral distribution in the core region. The contamination of nuclei heavier than helium is estimated smaller than 15\% at 1 PeV in all analyses.

In Figure~\ref{fig:pheknee} the direct measurements reported by CREAM \cite{cream2011} and NUCLEON \cite{nucleon2019} are also shown.
ARGO-YBJ is the only experiment that traced the (p+He) spectrum across the knee starting from an energy so low ($\approx$TeV) to overlap with direct measurements and to cross-calibrate the fluxes on a wide energy range (5--250 TeV).
These results show that, when indirect measurements are capable of selecting almost pure beams, their findings are in fair agreement with direct ones and confirm that current simulation models provide a satisfactory description of the EAS development in the atmosphere.
The cross-calibration of fluxes in this energy range, where the boundary line between `direct' and `indirect' measurements is uncertain, is very important.
The low energy threshold allowed also a calibration of the absolute energy scale at a level of 10\% exploiting the \emph{Moon Shadow} technique in the 1--30 TeV/Z range \cite{bartoli2011}.

As can be seen from the figure, the observations of Tibet AS$\gamma$ and ARGO-YBJ are in good agreement each other showing that the knee of the (p+He) energy spectrum is at $\approx$500--700 TeV, well below the energy of knee in the all-particle spectrum.
Similar conclusions have been obtained by the BASJE-MAX experiment located at 5200 m asl \cite{basje-mas} and by EAS-TOP at 2000 m asl \cite{aglietta2004} and by CASA-MIA at 1450 m asl \cite{casamia}.

\begin{figure}[H]
\includegraphics[width=0.8\textwidth]{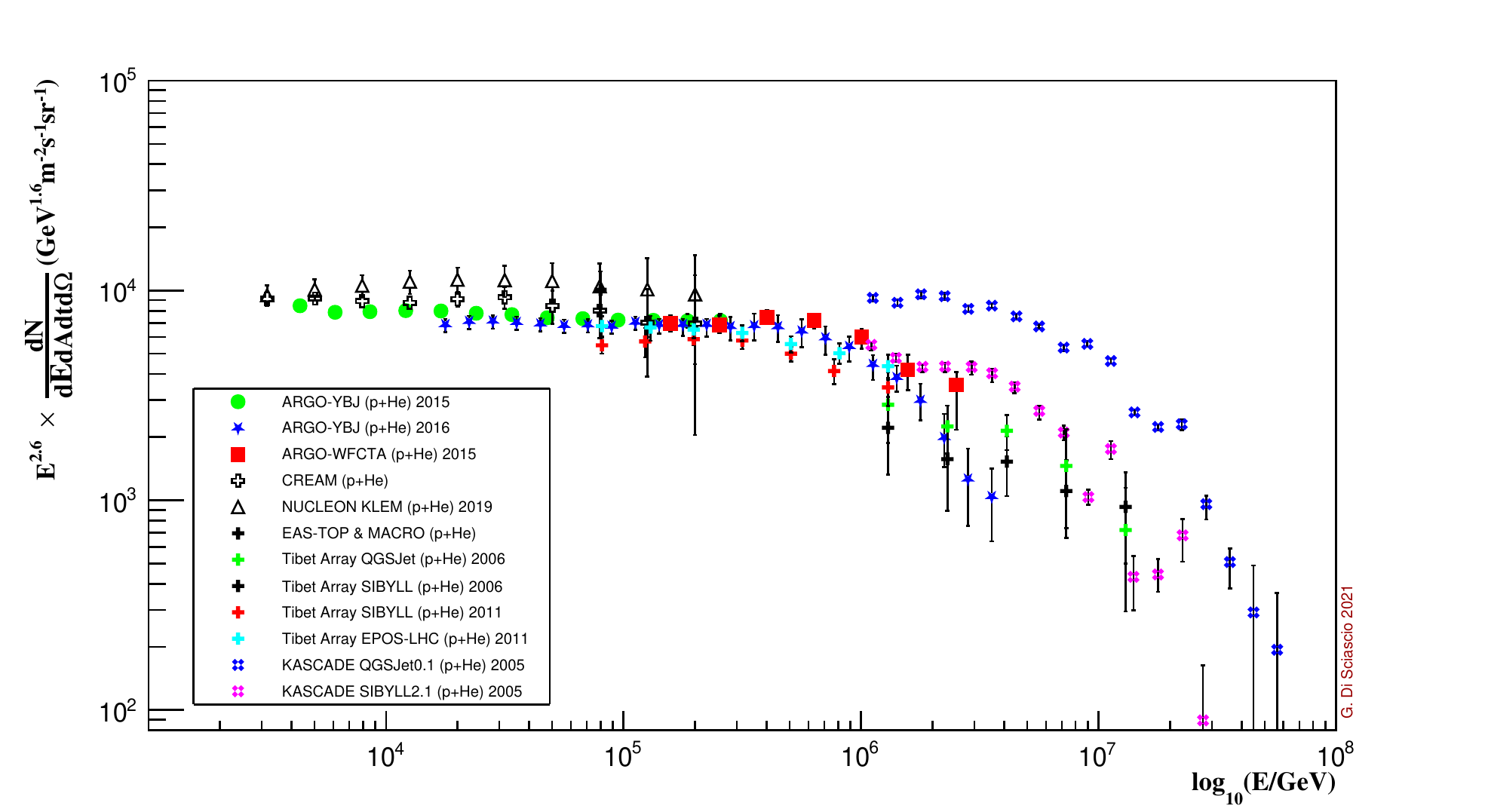}
\caption{The energy spectra shown in Figure~\ref{fig:pheknee} compared with the results obtained by the KASCADE experiment by using two different interaction models to interpret data \cite{kascade1,kascade2,kascade3}.
 \label{fig:phekascade}}%MDPI: please   change the hyphen to minus sign, -1 should be  −1
 \end{figure}

In Figure~\ref{fig:phekascade} the energy spectra of the light component reconstructed by the KASCADE experiment \cite{kascade1,kascade2,kascade3} with two different hadronic interaction models are added for comparison. The energy threshold is about 1 PeV and the experiment was located at sea level. KASCADE did use of a complex unfolding procedure to recover the elemental composition from the \emph{N}$_e-$\emph{N}$_{\mu}$ diagram in terms of 5 mass groups (p, He, CNO, MgSi, Fe).
As can be seen from the figure, both the spectra are at variance with the results obtained by Tibet AS$\gamma$ and ARGO-YBJ, suggesting that the knee of the CR all-particle spectrum at a few PeV is due to the bending of the light component. 
\vspace{-9pt}

%%%%%%%%%%%%%%%%%%%%%

%%%%%%%%%%%%%%%%%%%%%

\begin{figure}[H]
\includegraphics[width=0.8\textwidth]{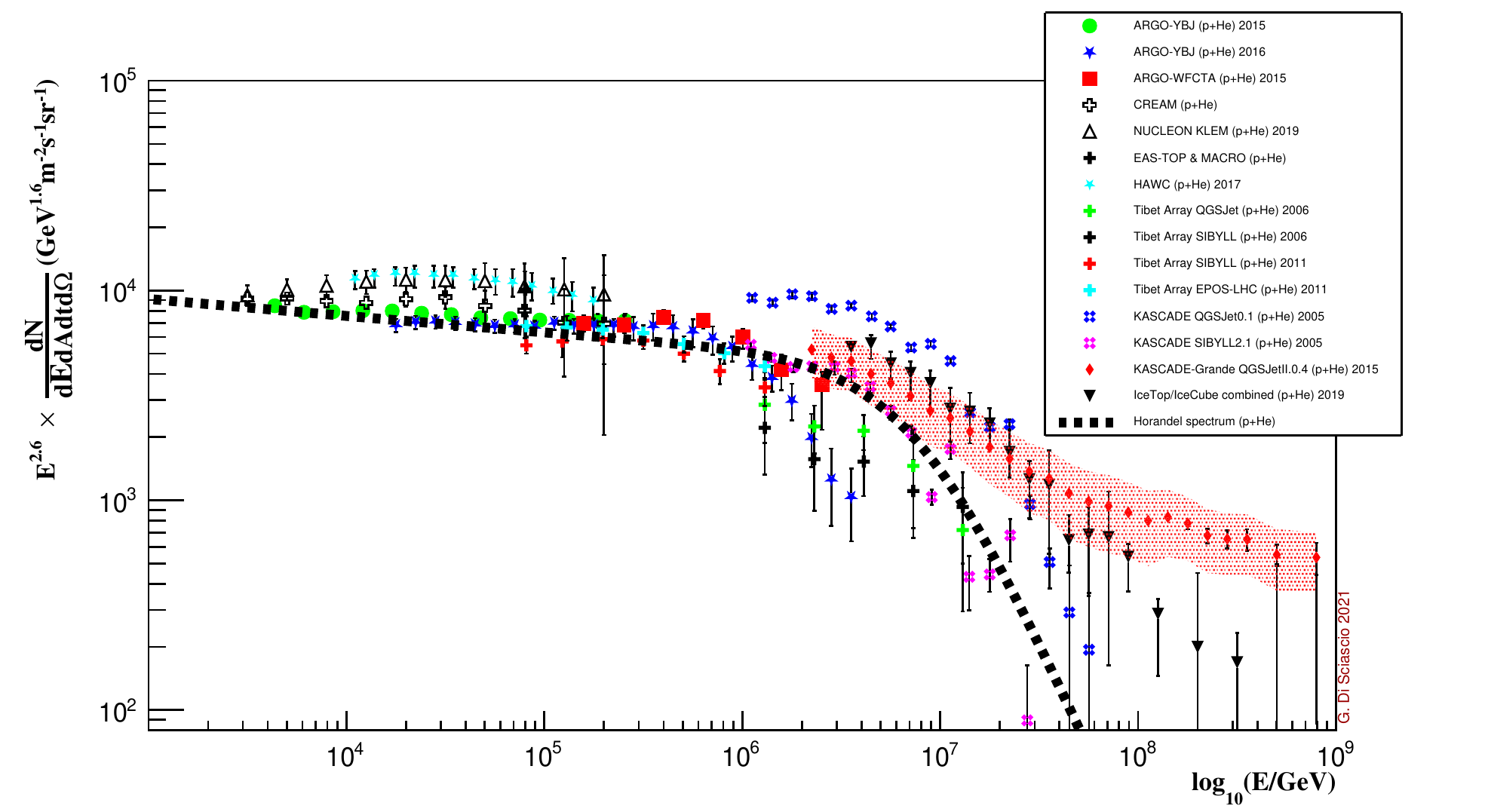}
\caption{The energy spectra shown in Figure~\ref{fig:phekascade} compared with the results obtained by HAWC in the 10–100 TeV region \cite{hawc} and by KASCADE–Grande \cite{kascade-grande} and IceTop/IceCube \cite{aartsen2019} combined above the PeV. The parametrization of the light component provided by H\"orandel \cite{horandel} is also shown.
 \label{fig:pheall}}%MDPI: please   change the hyphen to minus sign, -1 should be  −1
\end{figure}

\begin{figure}[H]
\includegraphics[width=0.99\textwidth]{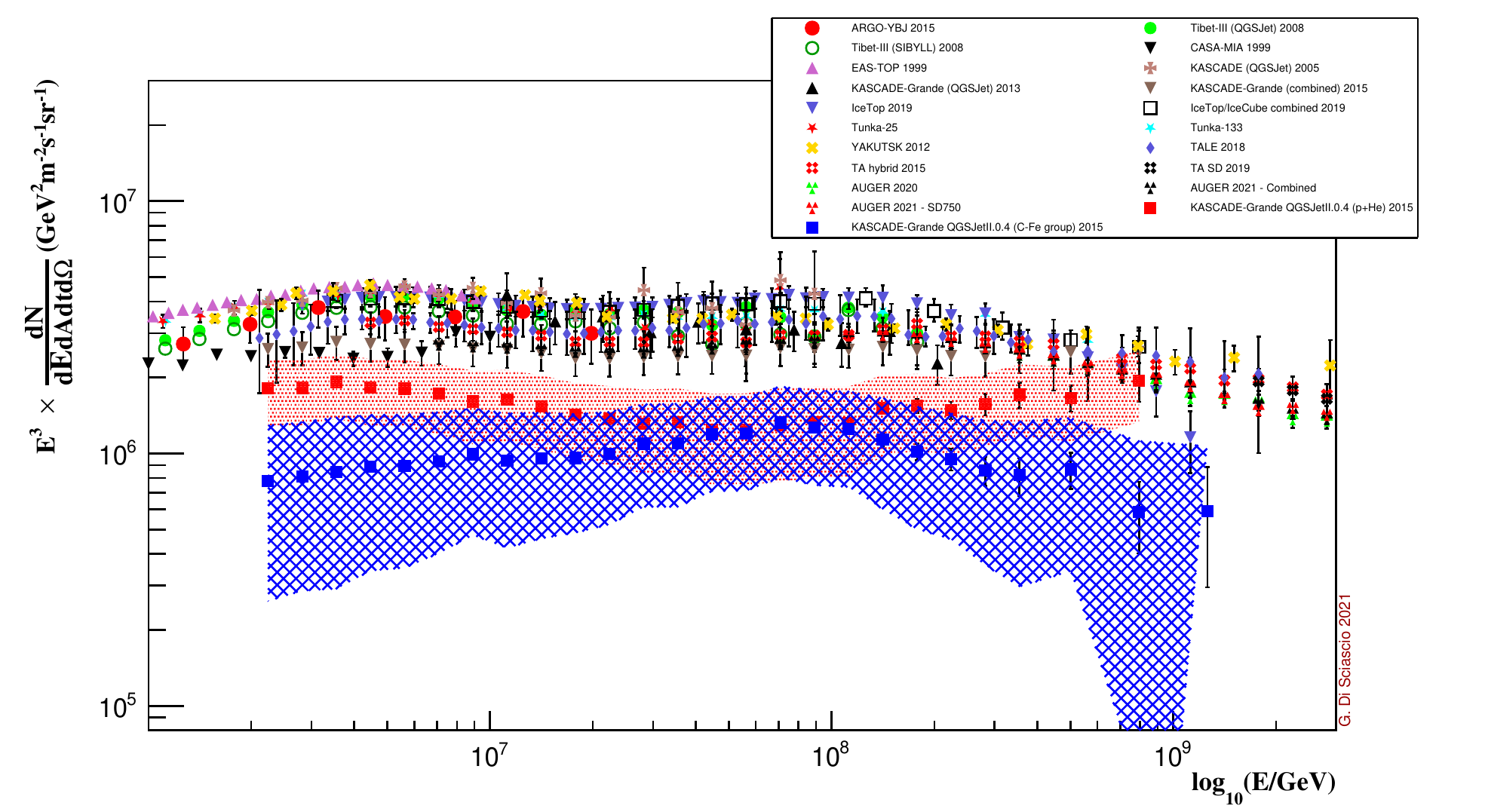}
\caption{The all–particle energy range in the 'transition' region measured by different experiments. The \emph{light} and \emph{heavy} components measured by KASCADE–Grande are also shown.
 \label{fig:transition}}%MDPI: please   change the hyphen to minus sign, -1 should be  −1
\end{figure}

All measurements of the light component up to about 10$^{18}$ eV are summarized in the Figure~\ref{fig:pheall} and compared with the parametrization provided by Horandel \cite{horandel}. 
Roughly speaking, we can separate the experiments that measured the (p+He) energy spectrum in the PeV range in 2 different groups
\begin{enumerate}
\item arrays located at extreme altitude (BASJE-MAS at 5200 m asl, ARGO-YBJ and Tibet AS$\gamma$ at 4300 m asl) observing a composition at the knee heavier than (p+He). These experiments did not exploit the measurement of the muon component to recover the elemental composition;
\item arrays located deeper in the atmosphere (KASCADE, KASCADE-Grande and IceTop/Icecube) reporting evidence that the light cut-off is located at a few PeV. In this case both low and high energy muons have been used in the classical study of $N$$_e$--$N$$_\mu$~correlation.
\end{enumerate}
\newpage
Measurements exploiting the longitudinal profile of the showers with Cherenkov detectors are conflicting too. The results obtained with the ARGO-YBJ hybrid detector (carpet and Cherenkov telescope) are in agreement with those of the carpet only, whereas the observations of Cherenkov light by Tunka-133 are consistent with KASCADE and KASCADE-Grande findings.
In Figure~\ref{fig:transition} the all-particle energy spectrum measured by several experiments in the energy region 10$^{16}$--10$^{18}$ eV is shown. As can be seen, the spectrum cannot be fitted by a single power law.
We observe a spectral hardening at $\sim$2 ~$\times$~ 10$^{16}$ eV and a steepening at $\sim$10$^{17}$ eV. 
This result was first pointed out by KASCADE-Grande experiment \cite{bertaina2011,bertaina2014}, then more firmly assessed with higher statistics and precision by Tunka-133 and IceTop-73~\cite{prosin2014,aartsen2019}, in particular for the feature at $\sim$2 ~$\times$~ 10$^{16}$ eV.

The \emph{light} (p+He) and \emph{heavy} (C-Fe group) components measured by KASCADE-Grande are also shown.
A knee is observed in the heavy component of CRs at \emph{E} = 10$^{16.92\pm 0.04}$ eV, which coincides within the uncertainties with the change of the slope in the all-particle energy spectrum around 10$^{17}$ eV. The spectral index changes from $-$2.76 $\pm$ 0.02 below the knee to $-$3.24 $\pm$ 0.05 above. 
At slightly higher energies (\emph{E} = 10$^{17.08\pm 0.09}$ eV), the light component shows a hardening of the slope, with the spectral index changing from \mbox{$-$3.25 $\pm$ 0.06} below this ankle-like structure to $-$2.79 $\pm$ 0.09 above. 
The positions of the changes of the slope as well as the intensities of the different components depend on the interaction model adopted to interpret the data. The knee in the heavy component seems visible also in the all-particle spectrum, as it is the dominant component.

The results obtained by Tunka-133, measuring the Cherenkov light deep in the atmosphere, suggest that the mass composition becomes heavier in the energy range \mbox{10$^{16}$--3~$\times$~10$^{16}$} eV, then stays heavy till 10$^{17}$ eV, where the composition starts becoming lighter.
IceTop/IceCube, exploiting the high-energy muons underground, indicates an increase of $\left<\ln A\right>$ in the energy range 10$^{16}$--3~$\times$~10$^{17}$ eV \cite{bertaina2014}.
However, as discussed in the previous sections, the average logarithmic mass of CR $\left<\ln A\right>$ is used to describe the evolution of the composition as a function of energy when the mass resolution of the experiments do not allow a discrimination between different mass groups.
%
%%%%%%%%%%%%%%%%%%%%%
%%%%%%%%%%%%%%%%%%%%%

In Figure~\ref{fig:transition} recent measurements of the all-particle energy spectrum down to about 100 PeV by the Pierre Auger Observatory are also reported. These observations suggest that the second knee is not a sharp feature, but a softening that extends in the interval 100--200 PeV \cite{auger2021}.
Preliminary results based on the distribution of the depth of the shower maximum are consistent with a spectrum dominated by heavy nuclei in the 10$^{17}$ eV range becoming lighter with increasing energy \cite{auger2018}.

The findings of Tunka-133, IceTop/IceCube and Auger are qualitatively in agreement with KASCADE-Grande.
Despite the large uncertainty in the absolute composition, a common general trend is reported, composition gets heavier through the knee region and becomes lighter approaching the ankle.

\begin{figure}[H]
\includegraphics[width=0.99\textwidth]{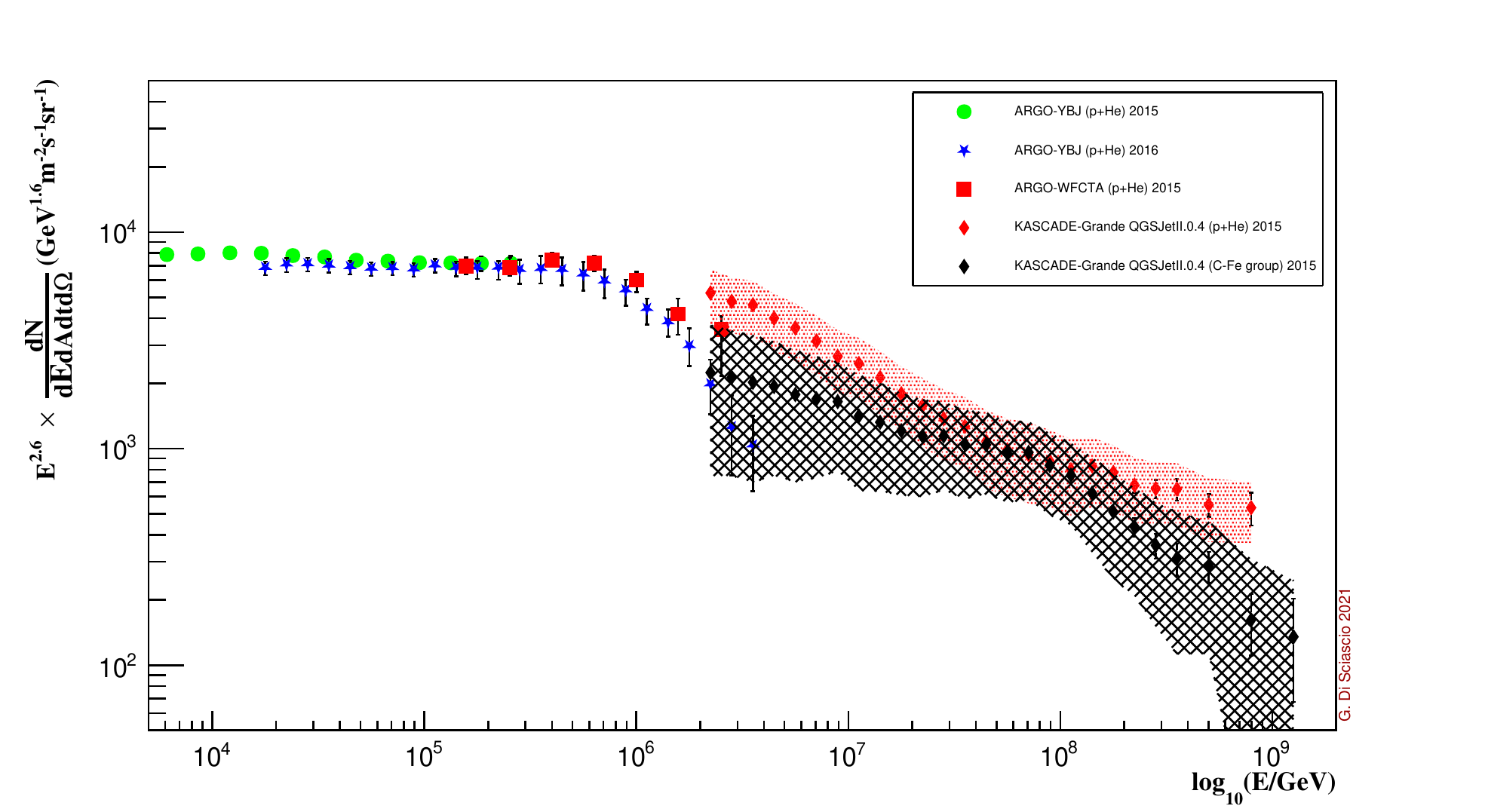}
\caption{The energy spectra of the light component measured by ARGO–YBJ compared to the \emph{light} and \emph{heavy} components measured by KASCADE–Grande.
 \label{fig:phe_argokg}}%MDPI: please   change the hyphen to minus sign, -1 should be  −1
\end{figure}

In conclusion, the observations of the different ground-based arrays show two conflicting results regarding the maximum energy at which the light component is accelerated in CR sources.
This disagreement is summarized in Figure~\ref{fig:phe_argokg} where the ARGO-YBJ results are compared with the KASCADE-Grande light and heavy spectra.
These results cannot be reconciled and show the existence of a still unknown systematic uncertainty that, as discussed in previous sections, could be due to the different array characteristics (altitude, coverage), the observables used (muons or shower core characteristics), and the dependence on hadronic interaction models. These are certainly among the major sources of systematic errors that affect the interpretation of shower array measurements (for a recent discussion see, for instance, refs. \cite{cazon2019a,cazon2019b}), although recent re-analyses of the KASCADE-Grande data with the latest versions of the post-LHC codes confirm previous results \cite{kascadeg2019}.

Important information could be deduced, in principle, by the measurement of the flux of atmospheric neutrinos, sensitive to the spectrum of parent CRs. In particular, the tail of the spectrum of atmospheric neutrinos is mainly shaped by the parent protons rather than by heavier element.
As a consequence, we expect different predictions for the flux of atmospheric neutrinos according to ARGO-YBJ and KASCADE proton energy spectra, predictions that, in principle, can be checked at  energies $E_{\nu}$ $\ge$ 100 TeV if the atmospheric neutrinos could be properly identified.
Unfortunately, in this energy region the total neutrino flux detected by IceCube departs from the existing predictions for atmospheric neutrinos suggesting the onset of an astrophysical component. 
The origin of such neutrinos is still unknown and current experimental uncertainties do not allow to draw clear conclusions \cite{mascaretti2020}.
%
%%%%%%%%%%%%%%%%%%%%%
%%%%%%%%%%%%%%%%%%%%%

%%%%%%%%%%%%%%%%%%%%%%%%%%%%%%%%%%%%%%%%%%

\section{What's Next}
\label{sec6}
The experimental situation in the 100 TeV--100 PeV energy region must be clarified to solve the longstanding problem of the origin of the knee and to give solid foundations to CR models up to the highest observed energies.
A new experiment, able to measure, at the right altitude and with high statistics, the elemental composition exploiting the techniques used so far in different apparatus, is mandatory to investigate the unknown uncertainties affecting the results so far obtained by shower arrays.

The only experiment that meets these requirement is LHAASO, a new multi-component array developed starting from the experience of the high altitude experiment ARGO-YBJ.
The apparatus is located at high altitude (4410 m asl, 600 g/cm$^2$) in the Daochen site, Sichuan province, P.R. China. 
LHAASO is expected to measure the energy spectrum, the elemental composition and the anisotropy of CRs in the energy range between 10$^{12}$ and 10$^{17}$~eV~\cite{cao2021nat,cao2021scie,lhaaso1,lhaaso-wb}.
The experiment is constituted by a 1 km$^2$ dense array of plastic scintillators and muon detectors. At the center of the array a 300~$\times$~300 m$^2$ water Cherenkov facility will allow the detection of TeV showers. An array of 18 wide field of view Cherenkov telescopes will image the longitudinal profile of events. Neutron monitors will study the hadronic component in the core of air showers.
LHAASO will study CR physics with different detectors and techniques starting from the TeV range, thus overlapping direct measurements in a wide interval.
In Tables \ref{tab:array-summary} and \ref{tab:array-summary-muon} the characteristics of the LHAASO-KM2A array are compared with other experiments. As can be seen, LHAASO will operate with a coverage of $\sim$0.5\% over a 1 km$^2$ area.
The sensitive area of muon detectors is unprecedented (more than 40,000 m$^2$), about 17 times larger than the CASA-MIA experiment, with a coverage of about 5\% over 1 km$^2$.
For the first time the $N$$_e$/$N$$_\mu$ correlation will be studied at high altitude with high statistics. 
This suite of independent instruments will also allow a deep investigation of the characteristics of the hadronic interaction models.
The capability of hybrid measurements with Cherenkov telescopes operated in combination with a shower array have been demonstrated by the ARGO-YBJ measurement of the light component energy spectrum.

In addition, LHAASO will act simultaneously as a wide aperture ($\sim$2 sr), continuously-operated gamma-ray telescope in the energy range between 10$^{11}$ and 10$^{15}$ eV. 
The first results obtained during the first year of data taking with only a portion of the apparatus opened for the first time the PeV sky to observations, showing that the Northern hemisphere contains a lot of galactic PeVatrons.

Other projects under way to investigate, with a much higher energy threshold, the high energy tail of the galactic spectrum and the transition region are HiSCORE \cite{hiscore} and GRAND \cite{grande}.

\section{Conclusions}

The results obtained by different experiments in the 10$^{14}$ to 10$^{18}$ eV region can be summarized as follows:
\begin{itemize}
\item {Knee energy region}%is the bold necessary?
\begin{enumerate}
\item All experiments observe an all-particle knee at $\approx$4~$\times$~10$^{15}$ eV.
\item The absolute fluxes are in good agreement with each other and with the direct measurements.
\item The elemental composition is conflicting. Experiments located at high and extreme altitude (BASJE-MAS, Tibet AS$\gamma$, ARGO-YBJ, EAS-TOP and CASA-MIA) reported evidence that the knee of the (p+He) component is below 1 PeV and that the composition at the all-particle knee energy is dominated by heavier nuclei. Experiments located deeper in atmosphere (KASCADE, KASCADE-Grande, IceTop/IceCube, Tunka-133) reported evidence that the proton knee is at the same energy of the all-particle knee.
\item A 10$^{-3}$--10$^{-4}$ Large Scale Anisotropy (LSA) amplitude is found at TeV energies~\cite{disciascio-anis}.
\item A 10$^{-4}$ Medium Scale Anisotropy (MSA) amplitude is observed at TeV energies~\cite{disciascio-anis}.
\end{enumerate}
\item {Transition region 10$^{16}$--10$^{18}$ eV}
\begin{enumerate}
\item The all-particle energy spectrum measured by different experiments are in good agreement within the systematics and with the measurements of UHE experiments.
\item A concave region is observed above 2 ~$\times$~ 10$^{16}$ eV with a steepening at $\sim$10$^{17}$ eV.
\item The dipole component of the LSA is smaller than 10$^{-2}$.
\end{enumerate}
\end{itemize}

The observed features in the all-particle energy spectrum seem to be consistent with the bending of different components in a rigidity-based scenario. However, rigidity models can~be 
\begin{itemize}
\item \emph{rigidity-acceleration} models, that is, the knee can be an acceleration feature, a source property, related to the maximum energy of particle acceleration inside the CR sources;
\item \emph{rigidity-confinement} models, that is, the knee is related to inefficient confinement of particles in the galaxy. In this case, the galaxy could contain \emph{'super-PeVatrons'}, astrophysical objects able to accelerate particles well beyond the PeV.
\end{itemize}

The first PeVatrons observed in the northern hemisphere by the LHAASO experiment show that SNRs are probably not the main sources of PeV CRs in our galaxy. The observation of sources emitting photons above the PeV in the North suggests the need of a wide field of view instrument in the Southern Hemisphere to monitor the Inner Galaxy and the Galactic Center looking for super-PeVatrons (SWGO \cite{swgo}, STACEX \cite{stacex2021}) to operate with CTA-South \cite{cta}.

In the coming years, the LHAASO experiment is expected to be able to measure the energy spectra of different mass groups up to 10$^{17}$ eV and to determine the energy of the proton knee, thus clarifying the origin of the knee in the all-particle spectrum. 
The suite of independent instruments that will be operated will also allow a deep study of the characteristics of the hadronic interaction models and to investigate the uncertainties related to the main techniques used to recover the elemental composition.

\vspace{6pt} 

%%%%%%%%%%%%%%%%%%%%%%%%%%%%%%%%%%%%%%%%%%
%% optional
%\supplementary{The following are available online at \linksupplementary{s1}, Figure S1: title, Table S1: title, Video S1: title.}


%%%%%%%%%%%%%%%%%%%%%%%%%%%%%%%%%%%%%%%%%%

\funding{This research received no external funding.}
%Please add: ``This research received no external funding'' or ``This research was funded by NAME OF FUNDER grant number XXX.'' and  and ``The APC was funded by XXX''. Check carefully that the details given are accurate and use the standard spelling of funding agency names at \url{https://search.crossref.org/funding}, any errors may affect your future funding.}

\institutionalreview{Not applicable.}
%In this section, please add the Institutional Review Board Statement and approval number for studies involving humans or animals. Please note that the Editorial Office might ask you for further information. Please add ``The study was conducted according to the guidelines of the Declaration of Helsinki, and approved by the Institutional Review Board (or Ethics Committee) of NAME OF INSTITUTE (protocol code XXX and date of approval).'' OR ``Ethical review and approval were waived for this study, due to REASON (please provide a detailed justification).'' OR ``Not applicable'' for studies not involving humans or animals. You might also choose to exclude this statement if the study did not involve humans or animals.}

\informedconsent{Not applicable.}
%Any research article describing a study involving humans should contain this statement. Please add ``Informed consent was obtained from all subjects involved in the study.'' OR ``Patient consent was waived due to REASON (please provide a detailed justification).'' OR ``Not applicable'' for studies not involving humans. You might also choose to exclude this statement if the study did not involve humans.

%Written informed consent for publication must be obtained from participating patients who can be identified (including by the patients themselves). Please state ``Written informed consent has been obtained from the patient(s) to publish this paper'' if applicable.}

\dataavailability{{Not applicable.}}
%In this section, please provide details regarding where data supporting reported results can be found, including links to publicly archived datasets analyzed or generated during the study. Please refer to suggested Data Availability Statements in section ``MDPI Research Data Policies'' at \url{https://www.mdpi.com/ethics}. You might choose to exclude this statement if the study did not report any data.} 

%\acknowledgments{The author thanks V. Verzi for useful discussions about experimental results in the transition region and for providing numerical tables of Auger energy spectra.
%This work is partly supported by the PIFI program of the Chinese Academy of Sciences (Grant No.113111WGZJTPYJY2020~\linebreak~0004) and by the National Natural Science Foundation of China (Grant No.U1931201).}

\conflictsofinterest{The authors declare no conflict of interest}
%Declare conflicts of interest or state ``The authors declare no conflict of interest.'' Authors must identify and declare any personal circumstances or interest that may be perceived as inappropriately influencing the representation or interpretation of reported research results. Any role of the funders in the design of the study; in the collection, analyses or interpretation of data; in the writing of the manuscript, or in the decision to publish the results must be declared in this section. If there is no role, please state ``The funders had no role in the design of the study; in the collection, analyses, or interpretation of data; in the writing of the manuscript, or in the decision to publish the~results''.} 

\begin{adjustwidth}{-\extralength}{0cm}
%\centering %% If there is a figure in wide page, please release command \centering
\reftitle{References}


%\externalbibliography{yes}
%\bibliography{your_external_BibTeX_file}

\begin{thebibliography}{999}

\bibitem[Battistoni&Grillo(1996)]{battistonigrillo1996}
Battistoni, G.; Grillo, A.F. Introduction to High-Energy Cosmic Ray Physics.      In Proceedings of the  ICTP School on Nonaccelerator Particle Astrophysics, Trieste, Italy, 17--28 July  1995; Preprint INFN/AE---96/05;  pp. 341--374.
\bibitem[DiSciascio(2019)]{disciascio2019}
Di Sciascio, G. Detection of Cosmic Rays from ground: An Introduction. {\em J. Phys. Conf. Ser.} {\bf 2019}, {\em 1263}, 012002. [\href{http://doi.org/10.1088/1742-6596/1263/1/012002}{CrossRef}]
\bibitem[Stanev(2014)]{stanev2014}
Stanev, T.C.R. Cosmogenic neutrinos and gamma rays. \emph{Physique} \textbf{2014}, 15, 349--356.  [\href{http://dx.doi.org/10.1016/j.crhy.2014.02.013}{CrossRef}]
\bibitem[Williams(1948)]{williams1948}
Williams, R.W. The structure of the large cosmic-ray air showers. {\em Phys. Rev.} {\bf 1948}, {\em 74}, 1689--1706.   [\href{http://dx.doi.org/10.1103/PhysRev.74.1689}{CrossRef}]
\bibitem[Linsley(1983)]{linsley1983}
Linsley, J. Spectra, anisotropies and composition of cosmic rays above 1000 GeV. \emph{Proc.  ICRC} \textbf{1983}, \emph{12}, 135L. 
\bibitem[khristiansen(1959)]{khristiansen1959}
Kulikov, G.V.; Khristiansen, G.B. On the size spectrum of extensive air showers. {\em Sov. Phys. JETP} {\bf 1959}, {\em 35}, 441--444. 

\bibitem[Miura(1962)]{miura1962}
Miura, I.; Hasegawa, H. Spectra of the Size and the Total Number of Mu-Mesons in EAS. {\em J. Phys. Soc. Jpn.} {\bf 1962}, {\em 17}, 84. 
\bibitem[Peters(1961)]{peters1961}
Peters, B. Primary cosmic radiation and extensive air showers. {\em Il Nuovo Cimento}  {\bf 1961}, {\em 22}, 800.  [\href{http://dx.doi.org/10.1007/BF02783106}{CrossRef}]
\bibitem[Drury(2012)]{drury12}
Drury, L.O. Origin of cosmic rays. {\em Astropart. Phys.} {\bf 2012}, {\em 39--40}, 52--60.  [\href{http://dx.doi.org/10.1016/j.astropartphys.2012.02.006}{CrossRef}]

\bibitem[Gabici(2016)]{gabici16}
Gabici, S.;  Gaggero, D.; Zandanel, F. Can supernova remnants accelerate protons up to PeV energies?   \emph{arXiv} {\bf 2016}, arXiv:1610.07638.  

\bibitem[Cao(2021)]{cao2021nat}
Cao, Z.; LHAASO Collaboration. Ultrahigh-energy photons up to 1.4 petaelectronvolts from 12 $\gamma$-ray Galactic sources. {\em Nature} {\bf 2021}, {\em 594}, 33--36.  [\href{http://dx.doi.org/10.1038/s41586-021-03498-z}{CrossRef}]
\bibitem[Cao(2021)]{cao2021scie}
Cao, Z.; LHAASO Collaboration. Peta–electron volt gamma-ray emission from the Crab Nebula. {\em Science} {\bf 2021}, {\em 373}, 425--430. 
\bibitem[Bartoli(2014)]{bartoli2014}
Bartoli, B.; Bernardini, P.; Bi, X.J.; Branchini, P.; Budano, A.; Camarri,  P.; Cao, Z.; Cardarelli, R.; Catalanotti, S.; Chen, S.Z.; et~al.  Identification of the TeV gamma-ray source ARGO J2031+4157 with the Cygnus Cocoon. {\em ApJ} {\bf 2014}, {\em 790},~152.  [\href{http://dx.doi.org/10.1088/0004-637X/790/2/152}{CrossRef}]
\bibitem[Aharonian(2019)]{aharonian2019}
Aharonian, F.; Yang, R.; de Ona Wilhelmi, E. Massive stars as major factories of Galactic cosmic rays. {\em Nat. Astron.} {\bf 2019}, {\em 3}, 561--567.  [\href{http://dx.doi.org/10.1038/s41550-019-0724-0}{CrossRef}]
\bibitem[Spurio(2018)]{spurio}
Spurio, M. \emph{Probes of Multimessenger Astrophysics}; Springer International Publishing: Cham, Switzerland, 2018. 
\bibitem[Gaisser(2016)]
{gaisser} Gaisser, T.K.; Engel, R.; Resconi, E. \emph{Cosmic Rays and Particle Physics}; Cambridge University Press: Cambridge, UK, 2016. 
\bibitem[Grieder(2010)]
{grieder} Grieder, P.K.F. \emph{Extensive Air Showers}; Springer International Publishing: Bern, Switzerland,  2010. 
\bibitem[Longair(1981)]
{longair} Longair, M.S. \emph{High Energy Astrophysics}; Cambridge University Press: Cambridge, UK, 1981. 
\bibitem[Blasi(2018)]
{blasi} Aloisio, R. \emph{Multiple Messengers and Challenges in Astroparticle Physics}; Springer International Publishing: Cham, Switzerland, 2018. 

\bibitem[Bartoli(2011)]{bartoli2011} 
Bartoli, B.; Bernardini, P.; Bi, X.J.;  Cao, Z.; Catalanotti, S.; Chen, S.Z.; Chen, T.L.; Cui, S.W.; Dai, B.Z.; D'Amone, A.; et~al.  Observation of the cosmic ray moon shadowing effect with the ARGO-YBJ experiment. {\em Phys. Rev. D} {\bf 2011}, {\em 84}, 022003. [\href{http://dx.doi.org/10.1103/PhysRevD.84.022003}{CrossRef}]
\bibitem[Ogio(2004)]{basje-mas}  
Ogio, S.; Kakimoto, F.; Kurashina, Y.; Burgoa, O.; Harada, D.; Tokuno, H.; Yoshii, H.; Morizawa, A.;  Gotoh, E.; Nakatani, H.; et~al. The energy spectrum and the chemical composition of primary cosmic rays with energies from 10$^{14}$ to 10$^{16}$ eV. {\em ApJ} \textbf{2004}, {\em 612}, 268.   [\href{http://dx.doi.org/10.1086/422510}{CrossRef}]
\bibitem[Antoni(2005)]{kascade1}
Antoni, T.; Apel, W.D.; Badea, A.F.; Bekk, K.; Bercuci A.; Blümer, J.; Bozdog, H.; Brancus, I.M.; Chilingarian, A.; Daumiller, K.;~et~al.  KASCADE measurements of energy spectra for elemental groups of cosmic rays: Results and open problems.  {\em Astropart. Phys.} \textbf{2005}, {\em 24}, 1.  [\href{http://dx.doi.org/10.1016/j.astropartphys.2005.04.001}{CrossRef}]

\bibitem[Apel(2009)]{kascade2}
Apel, W.D.; Arteaga, J.C.; Badea, A.F.; Bekk, K.; Blümer J.;Bozdog,  H.; Brancus, I.M.; Brüggemann, M.; Buchholz, P.; Cossavella, F.;~et~al. KASCADE Collaboration. Energy spectra of elemental groups of cosmic rays: Update on the KASCADE unfolding analysis. {\em Astropart. Phys.} \textbf{2009}, {\em 31}, 86. [\href{http://dx.doi.org/10.1016/j.astropartphys.2008.11.008}{CrossRef}]
%pls confirm OK

\bibitem[Apel(2013)]{kascade3}
Apel, W.D.; Arteaga-Velázquez,  J.C.; Bekk, K.; Bertaina, M.; Blümer, J.; Bozdog, H.; Brancus, I.M.; Cantoni, E.; Chiavassa, A.; Cossavella, F.; et~al.  KASCADE-Grande measurements of energy spectra for elemental groups of cosmic rays. {\em Astropart. Phys.} \textbf{2013}, {\em 47}, 54.   [\href{http://dx.doi.org/10.1016/j.astropartphys.2013.06.004}{CrossRef}]

\bibitem[Amenomori(2011)]{amenomori2011}
Amenomori, M.; Bi, X.J.; Chen, D.; Cui, S.W.; Danzengluobu; Ding, L.K.; Ding, X.H.; Fan, C.; Feng, C.F.; Fenget, Z.; et al.  Cosmic-ray energy spectrum around the knee obtained by the Tibet experiment and future prospects. {\em Adv. Space Res.} {\bf 2011}, {\em 47}, 629.   [\href{http://dx.doi.org/10.1016/j.asr.2010.08.029}{CrossRef}]

\bibitem[Glasmacher(1999)]{casamia}
Glasmacher, M.A.K.; Catanese, M.A.; Chantell M.C.; Covault, C.E.; Cronin, J.W.; Fick,  B.E.; Fortson, L.F.; Fowler, J.W.; Green, K.D.; Kieda, D.B.; et~al. CASA-MIA Collaboration. The cosmic ray composition between 10$^{14}$ and 10$^{16}$ eV. {\em Astropart. Phys.} {\bf 1999}, {\em 12}, 1.   [\href{http://dx.doi.org/10.1016/S0927-6505(99)00076-6}{CrossRef}]
\bibitem[Apel(2012)]{kascade-grande}
Apel, W.D.; Arteaga-Velázquez, J.C.; Bekk, K.; Bertainaet, M.;  Bluemer, J.; Bozdog,  H.; Brancus,  I.M.; Buchholz, P.; Cantoni, E.; Chiavassa, A.; et~al.  The spectrum of high-energy cosmic rays measured with KASCADE-Grande. {\em Astropart. Phys.} {\bf 2012}, {\em 36}, 183.   [\href{http://dx.doi.org/10.1016/j.astropartphys.2012.05.023}{CrossRef}]

\bibitem{prosin2014}
Prosin, V.V.; Berezhnev, S.F.; Budnev, N.M.; Brückner, M.; Chiavassa, A.; Chvalaev, O.A.; Dyachok, A.V.; Epimakhov, S.N.; Gafarov, A.V.; Gress, O.A.;  et~al. TUNKA Collaboration.  Results from Tunka-133 (5 years observation) and from the Tunka-HiSCORE prototype. In Proceedings of the 5th Roma International Conference on Astro-Particle physics (RICAP 14), Sicily, Italy, 30~September--3 October 2014.

\bibitem[Aartsen(2019)]{aartsen2019}
Aartsen, M.G.; Ackermann, M.; Adams, J.; Aguilar  J.A.; Ahlers, M.; Ahrens, M.; Alispach, C.; Andeen, K.; Anderson, T.; Ansseau,  I.; et~al. ICETOP Collaboration. Cosmic ray spectrum and composition from PeV to EeV using 3 years of data from IceTop and IceCube. {\em Phys. Rev. D} {\bf 2019}, {\em 100}, 082002.  [\href{http://dx.doi.org/10.1103/PhysRevD.100.082002}{CrossRef}]

\bibitem[Bartoli(2012)]{bartoli2012}
Bartoli, B.; Bernardini, P.;  Bi, X.J.; Bleve, C.; Bolognino I.; Branchini,  P.; Budano, A.; Calabrese Melcarne, A.K.; Camarri, P.; Cao, Z.;~et~al.  Light-component spectrum of the primary cosmic rays in the multi-TeV region measured by the ARGO-YBJ experiment. {\em Phys. Rev. D} {\bf 2012}, {\em 85}, 092005. [\href{http://dx.doi.org/10.1103/PhysRevD.85.092005}{CrossRef}]
%pls confirm OK

\bibitem[Amenomori(2006)]{amenomori2006}
Amenomori, M.; Ayabe, S.; Chen, D.; Cui, S.W.; Danzengluobu; Ding, L.K.; Ding, X.H.; Feng, C.F.; Feng, Z.Y.; Gao, X.Y.; et~al.  Are protons still dominant at the knee of the cosmic-ray energy spectrum? {\em Phys. Lett. B} {\bf 2006}, {\em 632}, 58--64.  [\href{http://dx.doi.org/10.1016/j.physletb.2005.10.048}{CrossRef}]

\bibitem[Heitler(1954)]{heitler}
Heitler, W. \emph{The Quantum Theory of Radiation}; Clarendon Press: Oxford, UK, 1944.
\bibitem[Matthews(2005)]{matthews}
Matthews, J. A Heitler model of extensive air showers. {\em Astropart. Phys.} \textbf{2005}, {\em 22}, 387.  [\href{http://dx.doi.org/10.1016/j.astropartphys.2004.09.003}{CrossRef}]


\bibitem[LetessierSelvon(2011)]{letessier2011}
Letessier-Selvon, A.; Stanev, T.  Ultrahigh energy cosmic rays. {\em Rev. Mod. Phys.} {\bf 2011}, {\em 83}, 907--942.  [\href{http://dx.doi.org/10.1103/RevModPhys.83.907}{CrossRef}]


\bibitem[Heck(1998)]{corsika}
Heck, D.; Knapp, J.; Capdevielle, J.N.; Schatz, G.; Thouw, T. \emph{CORSIKA: A Monte Carlo Code to Simulate Extensive Air Showers}; Forschungszentrum Karlsruhe GmbH: Karlsruhe, Germany, 1998.
%pls add journal name and volume and page 

\bibitem[Allison(2016)]{geant}
Allison, J.; Amako, K.; Apostolakis, J.; Arce, P.; Asai, M.; Aso, T.; Bagli, E.; Bagulya, A.; Banerjee, S.; Barrand, G.; et~al. Recent developments in Geant4. {\em NIM} {\bf 2016}, {\em A835}, 186--225.  [\href{http://dx.doi.org/10.1016/j.nima.2016.06.125}{CrossRef}]

\bibitem[Sciutto(2019)]{aires}
Sciutto, S.J. AIRES: A system for air shower simulations. \emph{arXiv} {\bf 2019}, arXiv:astro-ph/9911331. 

\bibitem[Horandel(2007)]{horandel2007}
Horandel, J.R. Cosmic Rays from the Knee to the Second Knee: 10$^{14}$ to 10$^{18}$ eV. {\em Mod. Phys. Lett. A} {\bf 2007}, {\em 22}, 1533--1551.  [\href{http://dx.doi.org/10.1142/S0217732307024139}{CrossRef}]

\bibitem[Kampert(2012)]{kampert2012}
Kampert, K.H.; Unger, M. Measurements of the cosmic ray composition with air shower experiments. {\em Astropart. Phys.} {\bf 2012}, {\em 35}, 660--678.  [\href{http://dx.doi.org/10.1016/j.astropartphys.2012.02.004}{CrossRef}]

\bibitem[Mollerach(2018)]{mollerach2018}
Mollerach, S.; Roulet, E. Progress in high-energy cosmic ray physics. {\em Progr. Part. Nucl. Phys.} {\bf 2018}, {\em 98}, 85--118.  [\href{http://dx.doi.org/10.1016/j.ppnp.2017.10.002}{CrossRef}]

\bibitem[Linsley(1977)]{linsley77}
Linsley, J. Structure of large air showers at depth 834 g/cm$^{-2}$: Applications. In Proceedings of the 15th International Cosmic Ray Conference, Plovdiv, Bulgaria, 13--26 August {1977}; Volume~12,  p.~89.

\bibitem[Fukui(1960)]{fukui1960}
Fukui, S.; Hasegawa, H.; Matano, T.;  Miura, I.; Oda, M.; Suga, K.; Tanahashi, G.; Tanaka, Y.  A study on the structure of the extensive air shower. {\em Suppl. Prog. Theor. Phys.} {\bf 1960}, {\em 16}, 1--53.  [\href{http://dx.doi.org/10.1143/PTPS.16.1}{CrossRef}]

\bibitem[Matano(1963)]{matano1963}
Matano, T.; Miura, I.; Nagano, M.; Oda, M.; Shibata, S.; Tanaka, Y.; Tanahashi, G.; Hasegawa, H. Extensive air showers---Studies of Tokyo group. In Proceedings of the 8th International Cosmic Ray Conference, Jaipur, India, 2--14 December {1963}; Volume~4,  p.~129.

\bibitem[Khristiansen(1963)]{khristiansen1963}
Vernov, S.N.; Khristiansen, G.B.; Abrosimov, A.M.; Atrashkevich, V.B.; Beliaeva, M.G. A descriptior of a modified complex installation for investigation of extensive air showers and new experimental data obtained by means of this installation. In Proceedings of the 8th International Cosmic Ray Conference, Jaipur, India, 2--14 December {1963}; Volume~4, p.~173 

\bibitem[Linsley(1962)]{linsley1962}
Linsley, J.; Scarsi, L.; Rossi, B. Energy spectrum and structure of large air showers. {\em J. Phys. Soc. Japan} {\bf 1962}, {\em 17}, 91 
\bibitem[Horandel(2008)]{horandel2008}
Horandel, J.R. Cosmic-ray composition and its relation to shock acceleration by supernova remnants. {\em Adv. Space Res.} {\bf 2008}, {\em 41}, 442--463. [\href{http://dx.doi.org/10.1016/j.asr.2007.06.008}{CrossRef}]
\bibitem[Lipari(2014)]{lipari2014}
Lipari, P. Cosmic rays and hadronic interactions. {\em C. R. Phys.} {\bf 2014}, {\em 15}, 357--366.  [\href{http://dx.doi.org/10.1016/j.crhy.2014.02.012}{CrossRef}]

\bibitem[Riehn(2020)]{riehn2020}
Riehn, F.; Engel, R.; Fedynitch, A.; Gaisser, T.K.; Stanev, T. Hadronic interaction model Sibyll 2.3d and extensive air showers. {\em Phys. Rev. D} {\bf 2020}, {\em 102}, 063002.  [\href{http://dx.doi.org/10.1103/PhysRevD.102.063002}{CrossRef}]

\bibitem[Kieda(2001)]{kieda2001}
Kieda, D.B.; Swordy, S.P.; Wakely, S.P.  A high resolution method for measuring cosmic ray composition beyond 10 TeV. {\em Astropart. Phys.} {\bf 2001}, {\em 15}, 287--303. [\href{http://dx.doi.org/10.1016/S0927-6505(00)00159-6}{CrossRef}]

\bibitem[Aharonian(2007)]{hess-fe2007}
Aharonian, F.; Akhperjanian, A.G.;  Bazer-Bachi, A.R.; Akhperjanian,  A.G.; Angüner, E.O.; Backes, M.; Balenderan, S.; Balzer, A.; Barnacka, A.; Becherini, Y.; et~al. HESS Collaboration. First ground-based measurement of atmospheric Cherenkov light from cosmic rays. {\em Phys. Rev. D} {\bf 2007}, {\em 75}, 042004.  [\href{http://dx.doi.org/10.1103/PhysRevD.75.042004}{CrossRef}]

\bibitem[Haungs(2003)]{haungs2003}
Haungs, A.; Rebel,  H.; Roth, M. Energy spectrum and mass composition of high-energy cosmic rays. {\em Rep. Prog. Phys.} {\bf 2003}, {\em 66}, 1145.  [\href{http://dx.doi.org/10.1088/0034-4885/66/7/202}{CrossRef}]
\bibitem[DiSciascio(2014)]{disciascio-rev}
Di Sciascio, G.  Main physics results of the ARGO-YBJ experiment. {\em Int. J. Mod. Phys. D} {\bf 2014}, {\em 23}, 1430019.  [\href{http://dx.doi.org/10.1142/S0218271814300195}{CrossRef}]

\bibitem[Aglietta(2004)]{aglietta2004}
Aglietta, M.; Alessandro, B.; Antonioli, P.; Arneodo, F.; Bergamasco, L.; Bertaina, M.; Castagnoli, C.; Castellina, A.; Chiavassa, A.; Cini Castagnoli, G.; et~al. EAS-TOP Collaboration. The cosmic ray primary composition in the “knee” region through the EAS electromagnetic and muon measurements at EAS-TOP. {\em Astropart. Phys.} {\bf 2004}, {\em 21}, 583.  [\href{http://dx.doi.org/10.1016/j.astropartphys.2004.04.005}{CrossRef}]

\bibitem[Garyaka(2007)]{garyaka2007}
Garyaka, A.P.; Martirosov, R.M.; Ter-Antonyan, S.V.; Nikolskaya, N.; Gallant, Y.A.; Jones, L.; Procureur, J. GAMMA Collaboration. Rigidity-dependent cosmic ray energy spectra in the knee region obtained with the GAMMA experiment. {\em Astropart. Phys.} {\bf 2007}, {\em 28}, 169.  [\href{http://dx.doi.org/10.1016/j.astropartphys.2007.04.004}{CrossRef}]

\bibitem[Tanaka(2012)]{tanaka2012}
Tanaka, H.; Dugad, S.R.; Gupta, S.K.; Jain A.; Mohanty, P.K.; Rao, B.S.; Ravindran, K.C.; Sivaprasad, K.; Tonwar, S.C.; Hayashi, Y.;~et~al. GRAPES Collaboration. Studies of the energy spectrum and composition of the primary cosmic rays at 100–1000 TeV from the GRAPES-3 experiment. {\em J. Phys. G Nucl. Part. Phys.} {\bf 2012}, {\em 39}, 025201.  [\href{http://dx.doi.org/10.1088/0954-3899/39/2/025201}{CrossRef}]

\bibitem[Bartoli(2015)]{argo-hybrid2}
Bartoli, B.; Bernardini, P.; Bi,  X.J.; Cao, Z.; Catalanotti, S.; Camarri, P.; Cao, Z.; Cardarelli, R.; Catalanotti, S.; Chen, S.Z.; et al.  Knee of the cosmic hydrogen and helium spectrum below 1 PeV measured by ARGO-YBJ and a Cherenkov telescope of LHAASO. {\em Phys. Rev. D} {\bf 2015}, {\em 92}, 092005. [\href{http://dx.doi.org/10.1103/PhysRevD.92.092005}{CrossRef}]
%pls confirm OK

\bibitem[Yoon(2011)]{cream2011}
Yoon, Y.S.; Ahn, H.S.;  Allison, P.S.; Bagliesi, M.G.;  Beatty, J.; Bigongiari, G.; Boyle, P.J.; Childers, J.T.; Conklin, N.B.; Coutu, S.; et al. CREAM Collaboration. Cosmic-ray proton and helium spectra from the first CREAM flight. {\em  ApJ} {\bf 2011}, {\em 728}, 122.  [\href{http://dx.doi.org/10.1088/0004-637X/728/2/122}{CrossRef}]

\bibitem{nucleon2019}
Grebenyuk, V.; Karmanov, D.; Kovalev, I.; Kovalev, I.; Kudryashov, I.; Kurganov, A.; Panov, A.; Podorozhny, D.; Porokhovoy, S.; Sveshnikova, L.; et~al. NUCLEON Collaboration. Energy spectra of abundant cosmic-ray nuclei in the NUCLEON experiment. {\em Adv. Space Res.} {\bf 2019}, {\em 64}, 2546.  [\href{http://dx.doi.org/10.1016/j.asr.2019.10.004}{CrossRef}]

\bibitem[Bartoli(2015)]{bartoli15}
Bartoli, B.; Bernardini, P.; Bi, X.J.; Cao, Z.; Catalanotti, S.;  Chen, S.Z.; Chen, T.L.; Cui, S.W.; Dai, B.Z.; D'Amone, A.; et~al. Cosmic ray proton plus helium energy spectrum measured by the ARGO-YBJ experiment in the energy range 3–300 TeV. {\em Phys. Rev. D} {\bf 2015}, {\em 91}, 112017.  [\href{http://dx.doi.org/10.1103/PhysRevD.91.112017}{CrossRef}]

\bibitem[Bartoli(2015)]{argo-bigpad}
Bartoli, B.; Bernardini, P.; Bi, X.J.; Branchini, P.; Budano, A.;  Chen, S.Z.; Chen, T.L.; Cui, S.W.; Dai, B.Z.; D'Amone, A.; et~al.  The analog Resistive Plate Chamber detector of the ARGO-YBJ experiment. {\em Astropart. Phys.} {\bf 2015}, {\em 67}, 47.  [\href{http://dx.doi.org/10.1016/j.astropartphys.2015.01.007}{CrossRef}]

\bibitem[Bartoli(2014)]{argo-hybrid1}
Bartoli, B.; Bernardini, P.; Bi, X.J.; Bolognino, I.; Branchini, P.; Budano, A.; Calabrese Melcarne, A.K.; Camarri, P.; Cao, Z.; Cardarelli, R.; et~al. Energy spectrum of cosmic protons and helium nuclei by a hybrid measurement at 4300 m asl. {\em Chinese Phys.} {\bf 2014}, {\em  C38}, 045001.  [\href{http://dx.doi.org/10.1088/1674-1137/38/4/045001}{CrossRef}]

\bibitem[Horandel(2003)]{horandel}
H\"orandel, J.H. On the knee in the energy spectrum of cosmic rays. {\em Astropart. Phys.} {\bf 2003}, {\em 19}, 193.  [\href{http://dx.doi.org/10.1016/S0927-6505(02)00198-6}{CrossRef}]

\bibitem[Alfaro(2017)]{hawc}
Alfaro, R.; Alvarez, C.; Álvarez, J.D.;  Arceo, R.; Avila Rojas, D.; Ayala Solares, H.A.; Barber, A.S.; Becerril, A.; Belmont-Moreno, E.; BenZvi, S.Y.; et~al. HAWC Collaboration. All-particle cosmic ray energy spectrum measured by the HAWC experiment from 10 to 500 TeV. {\em Phys. Rev. D} {\bf 2017}, {\em 96}, 122001.  [\href{http://dx.doi.org/10.1103/PhysRevD.96.122001}{CrossRef}]

\bibitem[Bertaina(2011)]{bertaina2011}
Bertaina, M.E.;  Apel, W.D.; Hörandel, J.R.; Wommer, M.; Blumer, J.; Bozdog, H.; Brancus, I.M.;
Buchholz, P.; Cantoni, E.; Chiavassa, A.; et~al. KASCADE-Grande Collaboration, The cosmic ray energy spectrum in the range 10$^{16}$–10$^{18}$ eV measured by KASCADE-Grande. {\em Astrophys. Space Sci. Trans.} {\bf 2011}, {\em 7}, 229.  [\href{http://dx.doi.org/10.5194/astra-7-229-2011}{CrossRef}]

\bibitem[Bertaina(2014)]{bertaina2014}
Bertaina, M.E. Cosmic rays from the knee to the ankle. {\em C. R. Phys.} {\bf 2014}, {\em 15}, 300--308.  [\href{http://dx.doi.org/10.1016/j.crhy.2014.03.001}{CrossRef}]

\bibitem[Abreu(2021)]{auger2021}
Abreu, P.; Aglietta, M.; Albury, J.M.; Almela, A.; Alvarez-Muñiz, J.; Alves Batista, R.; Anastasi, G.A.; Anchordoqui, L.; Andrada, B.; Allekotte, I.; et al. The energy spectrum of cosmic rays beyond the turn-down around 10$^{17}$ eV as measured with the surface detector of the Pierre Auger Observatory. {\em Eur. Phys. J.} {\bf 2021}, {\em 81}, 966.  [\href{http://dx.doi.org/10.1140/epjc/s10052-021-09700-w}{CrossRef}]

\bibitem[Bellido(2018)]{auger2018}
Bellido, J.;  Aglietta, M.; Albury, J.M.; Allekotte, I.; Almeida Cheminant, K.; Almela, A.; Alvarez-Muñiz, J.; Alves Batista, R.; Anastasi, G.A.; Anchordoqui, L.; et al. (The Pierre Auger Collaboration). Depth of maximum of air-shower profiles at the Pierre Auger Observatory: Measurements above 10$^{17.2}$ eV and Composition Implications. In Proceedings of the 35th International Cosmic Ray Conference (ICRC2017),  Busan, Korea, 10–20 July 2017.

\bibitem[Cazon(2019a)]{cazon2019a}
Cazon, L. Probing High-Energy Hadronic Interactions with Extensive Air Showers. In Proceedings of the  36th International Cosmic Ray Conference (ICRC2019), Madison, WI, USA, 24 July–1 August 2019; Volume 358, p. 5. 

\bibitem[Cazon(2019b)]{cazon2019b}
Cazon, L. Working Group Report on the Combined Analysis of Muon Density Measurements from Eight Air Shower Experiments. In Proceedings of the 36th International Cosmic Ray Conference (ICRC2019), Madison, WI, USA, 24 July–1 August 2019; Volume~358, p. 214. 

\bibitem[Kang(2019)]{kascadeg2019}
Kang, D.; Haungs, A.; Apel, W.D.; Arteaga-Velázquez, J.C.; Beket, K.; Bertaina, M.; Blümer, J.; Bozdog, H.; Cantoni, E.; Chiavassa, A.; et~al.   Latest Results from the KASCADE-Grande Data Analysis. In Proceedings of the 36th International Cosmic Ray Conference (ICRC2019), Madison, WI, USA, 24 July–1 August 2019; Volume~ 358, p. 306. 

\bibitem{mascaretti2020}
Mascaretti, C.; Blasi, P.; Evoli, C. Atmospheric neutrinos and the knee of the cosmic ray spectrum. {\em Astropart. Phys.} {\bf 2020}, {\em 114}, 22--29.  [\href{http://dx.doi.org/10.1016/j.astropartphys.2019.06.002}{CrossRef}]

\bibitem[DiSciascio(2016)]{lhaaso1}
Di Sciascio, G.;  LHAASO Collaboration.  The LHAASO experiment: From Gamma-Ray Astronomy to Cosmic Rays.
{\em Nucl.  Part. Phys. Proc.} {\bf 2016}, {\em 279–281}, 166--173.  [\href{http://dx.doi.org/10.1016/j.nuclphysbps.2016.10.024}{CrossRef}]

\bibitem[Bai(2019)]{lhaaso-wb}
Bai, X.;  Bi, B.Y.; Bi, X.J.; Cao, Z.; Chen, S.Z.;  Chen, Y.; Chiavassa, A.; Cui, X.H.; Dai, Z.G.; della Volpe, D.; et~al.  The Large High Altitude Air Shower Observatory (LHAASO) Science White Paper. \emph{arXiv} \textbf{2019}, arXiv:1905.02773. 

\bibitem[Tluczykont(2014)]{hiscore}
Tluczykont, M.;  Hampf, D.; Horns, D.; Spitschan, D.; Kuzmichev, L.; Prosin, V.; Spiering, C.; Wischnewski, R.; HiSCORE Collaboration. The HiSCORE concept for gamma-ray and cosmic-ray astrophysics beyond 10 TeV.  {\em Astropart. Phys.} {\bf 2014}, {\em 56}, 42--53.  [\href{http://dx.doi.org/10.1016/j.astropartphys.2014.03.004}{CrossRef}]

\bibitem[Alvarez-Muniz(2020)]{grande}
Alvarez-Muniz, J.; Batista, R.A.; Bolmont,  J.; Bolmont, J.; Bustamante, M.; Carvalho, W., Jr.; Charrier, D.; Cognard, I.; Decoene, V.; Denton, P.B.;  et al. The giant radio array for neutrino detection (GRAND): Science and design. {\em Sci. China} {\bf 2020}, {\em 63}, 219501.  [\href{http://dx.doi.org/10.1007/s11433-018-9385-7}{CrossRef}]

\bibitem[DiSciascio(2013)]{disciascio-anis}
Di Sciascio, G.; Iuppa, R.  \emph{Homage to the Discovery of Cosmic Rays}; Perez-Peraza, J.A., Ed.;  Nova Science Publishers: New York, NY, USA, 2013; Chapter 9, pp. 221--257. 

\bibitem[Albert(2019)]{swgo}
Albert A.;  Alfaro, R.; Ashkar, H.; Alvarez, C.; Álvarez, J.; Arteaga-Velázquez, J.C.; Ayala Solares, H.A.; Arceo, R.; Bellido, J.A.; BenZvi, S.; et~al.  Science Case for a Wide Field-of-View Very-High-Energy Gamma-Ray Observatory in the Southern Hemisphere. \emph{arXiv} {\bf 2019}, arXiv:1902.08429v1.

\bibitem[Rodriguez-Fernandez(2021)]{stacex2021}
Rodriguez-Fernandez, G.; Bigonciari, C.; Bulgarelli, A.; Camarri, P.; Cardillo, M.; Di Sciascio, G.; Valentina, F.; Marco, R.; Giovanni, P.; Rinaldo, S.; et~al.  STACEX: A RPC-based detector for a multi-messenger Southern observatory in the GeV-PeV range. In Proceedings of the 37th International Cosmic Ray Conference (ICRC2021),  Berlin, Germany, 12–23 July 2021.

\bibitem[Acharya(2017)]{cta}
Acero, F.; Acharya, B.S.; Acín Portella, V.; Adams, C.; Agudo, I.; Aharonian, F.; Al Samarai, I.; Alberdi, A.; Alcubierre, M.; Alfaro, R.; et al.  Cherenkov Telescope Array. In Proceedings of the 35th International Cosmic Ray Conference (ICRC2017),   Busan, Korea, 10–20 July 2017.

%
%
%ref 80 have no citation, pls revise
\end{thebibliography}

%=====================================
% References, variant B: internal bibliography
%=====================================

\end{adjustwidth}

% If authors have biography, please use the format below
%\section*{Short Biography of Authors}
%\bio
%{\raisebox{-0.35cm}{\includegraphics[width=3.5cm,height=5.3cm,clip,keepaspectratio]{Definitions/author1.pdf}}}
%{\textbf{Firstname Lastname} Biography of first author}
%
%\bio
%{\raisebox{-0.35cm}{\includegraphics[width=3.5cm,height=5.3cm,clip,keepaspectratio]{Definitions/author2.jpg}}}
%{\textbf{Firstname Lastname} Biography of second author}

% The following MDPI journals use author-date citation: Admsci,  Arts, Econometrics, Economies, Genealogy, Humanities, IJFS, Jintelligence, JRFM, Languages, Laws, Literature, Religions, Risks, Social Sciences. For those journals, please follow the formatting guidelines on http://www.mdpi.com/authors/references
% To cite two works by the same author: \citeauthor{ref-journal-1a} (\citeyear{ref-journal-1a}, \citeyear{ref-journal-1b}). This produces: Whittaker (1967, 1975)
% To cite two works by the same author with specific pages: \citeauthor{ref-journal-3a} (\citeyear{ref-journal-3a}, p. 328; \citeyear{ref-journal-3b}, p.475). This produces: Wong (1999, p. 328; 2000, p. 475)

%%%%%%%%%%%%%%%%%%%%%%%%%%%%%%%%%%%%%%%%%%
%% for journal Sci
%\reviewreports{\\
%Reviewer 1 comments and authors’ response\\
%Reviewer 2 comments and authors’ response\\
%Reviewer 3 comments and authors’ response
%}
%%%%%%%%%%%%%%%%%%%%%%%%%%%%%%%%%%%%%%%%%%
\end{document}